\documentclass[acmsmall,screen]{acmart} 
\usepackage{graphicx}
\usepackage{subcaption}
\usepackage{adjustbox}
\usepackage{tabularx}
\usepackage{multirow} 
\usepackage{changepage}
\usepackage{balance}
\usepackage[super]{nth}
\usepackage{amsmath, amsfonts}
\usepackage{algorithmic}
\usepackage{booktabs}
\usepackage{array}
\usepackage{textcomp}
\usepackage{url}
\usepackage{CJKutf8}

\usepackage[most]{tcolorbox}
\definecolor{custom-gray}{cmyk}{0, 0, 0, 0.7, 1.00}
\newtcbtheorem[no counter]{Summary}{\hskip-0.97em}{enhanced,drop shadow={black!50!white},
 coltitle=white,
 top=0.15in,
 attach boxed title to top left=
 {xshift=1.5em,yshift=-\tcboxedtitleheight/2},
 boxed title style={size=small,colback=custom-gray}
}{summary}

\usepackage{listings}
\usepackage{xcolor}
\usepackage[super]{nth}
\definecolor{codegreen}{rgb}{0,0.6,0}
\definecolor{codegray}{rgb}{0.5,0.5,0.5}
\definecolor{codepurple}{rgb}{0.58,0,0.82}
\definecolor{backcolour}
{rgb}{0.95,0.95,0.92}

\lstdefinestyle{mystyle}{
    backgroundcolor=\color{backcolour},   
    commentstyle=\color{codegreen},
    keywordstyle=\color{magenta},
    numberstyle=\tiny\color{codegray},
    stringstyle=\color{codepurple},
    basicstyle=\ttfamily\footnotesize,
    breakatwhitespace=false,breaklines=true,captionpos=b,keepspaces=true,numbers=left,numbersep=5pt,showspaces=false,showstringspaces=false,showtabs=false,tabsize=2
}
\newcommand{\rqone}{RQ1: What is the prevalence of SATD in ML-based systems?}
\newcommand{\rqtwo}{RQ2: What are the different types of SATD in ML-based systems?}
\newcommand{\rqthree}{RQ3: Which stages of the ML pipeline are more prone to SATD?}
\newcommand{\rqfour}{RQ4: How long does SATD survive in ML-based systems?}
\newcommand{\rqfive}{RQ5: What are the characteristics of long-lasting SATDs in ML-based systems?}

\def\BibTeX{{\rm B\kern-.05em{\sc i\kern-.025em b}\kern-.08em
    T\kern-.1667em\lower.7ex\hbox{E}\kern-.125emX}}

\AtBeginDocument{%
  \providecommand\BibTeX{{%
    \normalfont B\kern-0.5em{\scshape i\kern-0.25em b}\kern-0.8em\TeX}}}

\setcopyright{acmcopyright}

\acmJournal{TOSEM}

\begin{document}
\title{An Empirical Study of Self-Admitted Technical Debt in Machine Learning Software}
	
\author{Aaditya Bhatia}
\email{aaditya.bhatia@queensu.ca}
\affiliation{%
  \institution{SAIL, Queen's University}
  \city{Kingston}
  \country{Canada}
}

\author{Foutse Khomh}
\email{foutse.khomh@polymtl.ca}
\affiliation{%
  \institution{Polytechnique Montréal}
  \country{Canada}
}

\author{Bram Adams}
\email{bram.adams@queensu.ca}
\affiliation{%
  \institution{MCIS, Queen's University}
  \city{Kingston}
  \country{Canada}
}

\author{Ahmed E Hassan}
\email{ahmed@cs.queensu.ca}
\affiliation{%
  \institution{SAIL, Queen's University}
  \city{Kingston}
  \country{Canada}
}

\renewcommand{\shortauthors}{Bhatia, et al.}

\begin{abstract}
The emergence of open-source ML libraries such as TensorFlow and Google Auto ML has enabled developers to harness state-of-the-art ML algorithms with minimal overhead. However, during this accelerated ML development process, said developers may often make sub-optimal design and implementation decisions, leading to the introduction of technical debt that, if not addressed promptly, can significantly impact on the quality of ML-based software. Developers frequently acknowledge these sub-optimal design and development choices through code comments written during development. These comments, which often highlight areas requiring additional work or refinement in the future are known as \textit{self-admitted technical debt (SATD)}. While prior research has demonstrated that SATD can serve as a reliable indicator of technical debt and has extensively studied SATD in traditional (non-ML) software, little attention has been given to this issue in the context of ML. This paper aims to investigate the occurrence of SATD in ML code by analyzing 318 open-source ML projects across five domains, along with 318 non-ML projects. We detected SATD in source code comments in various snapshots of the studied projects, conducted a manual analysis of a sample of the identified SATD to comprehend the nature of technical debt in the ML code, and performed a survival analysis of the SATD to understand the evolution dynamics of such debts. Our analyses yielded the following observations: (i) Machine learning projects have a median percentage of SATD that is twice that of non-machine learning projects. (ii) ML pipeline stages for \textit{data preprocessing} and \textit{model generation logic} are more susceptible to debt than \textit{model validation} and \textit{deployment} stages. (iii) SATDs appear in ML projects earlier in the development process compared to non-ML projects. (iv) Long-lasting SATDs are typically introduced during extensive code changes that span multiple files, which exhibit low complexity.

Our research contributes to the understanding of technical debt in an ML context and underscores the need for targeted debt management strategies. This contribution is particularly relevant for developers and stakeholders in ML projects by aiding them in identifying and addressing technical debt proactively and paving the way for future research in developing automated tools and methodologies for managing SATD in an ML environment.
\end{abstract}

\keywords{Self Admitted Technical Debt, Machine Learning, Temporal Analysis}
\maketitle

\section{Introduction}\label{sec.intro}

The burgeoning of the Machine Learning (ML) era signals increased deployment of ML software in a wide range of software domains including large scale systems for healthcare, mobility, banking, transportation, policing, marketing, and social media~\cite{jordan2015machine}. ML-based systems, i.e., projects that integrate ML models with other non-ML components, serve as backbone for this expansive integration across domains. 

Unless they reuse third-party models, ML-based systems involve the training of custom ML models, which is achieved using a ML pipeline composed of multiple stages like data preprocessing, ML model training, model evaluation, and deployment~\cite{bhatia2023towards}. The development of ML-based systems necessitates continuous management of ML-related assets~\cite{idowu2021asset} (e.g., training data, ML models, job execution data) and introduces additional challenges that are not prevalent in traditional (non-ML) software~\cite{sato2019continuous}. 
While popular ML toolkits such as Scikit-Learn, TensorFlow, and Auto ML, have enabled developers to easily leverage state-of-the-art ML algorithms, the specification of ML algorithms is far from trivial~\cite{d2022underspecification}, and development and maintenance practices for software built around ML differ from those of traditional software products~\cite{wan2019does}. For instance, the dependency of ML code on external data makes it difficult to enforce strict abstraction boundaries~\cite{Sculley}. Apart from ML toolkit frameworks, ML-based systems also need (amongst others) established protocols for iterative model training, maintaining a structured log of experiments and their outcomes, while pruning dead code paths to promptly maintain codebase clarity.

As a result of these challenges and the tight deadlines for rushing AI products to the market, developers and data scientists can make suboptimal decisions for short-term benefits~\cite{obrien202223} when working on their code, models and data, a phenomenon described by Cunningham~\cite{cunningham1992wycash} as technical debt (TD). In software engineering, TD manifests as 1) optimization issues where optimization for short-term puts long-term into technical (and economic) jeopardy, 2) short-cuts in software implementation giving the perception of success until their consequences have crippling effects, and 3) development and architectural decisions prioritizing short-term causing long-term consequences~\cite{brown2010managing}. TD may represent inferior development implementations in software design, missing/incomplete documentation, known defects, code, testing, or software requirements as identified as Bavota et al.~\cite{bavota2016large} for traditional software; or as data dependencies, hidden feedback loops or pipeline jungles for ML software as identified by Sculley et al.~\cite{Sculley}. TD results from decisions made while evaluating competing development tradeoffs; these decisions, just like financial debt, will require correction in the future~\cite{klinger2011enterprise}. Such a correction may come with non-trivial interest, for instance, as identified by Guo et al.~\cite{guo2016exploring}, it can take up to 100\% interest probability, i.e., taking as many hours to pay off the TD as the principal time taken to develop the initial software artifact.

\begin{figure}[!t]
  \centering
  \includegraphics[width=0.7\columnwidth,keepaspectratio]{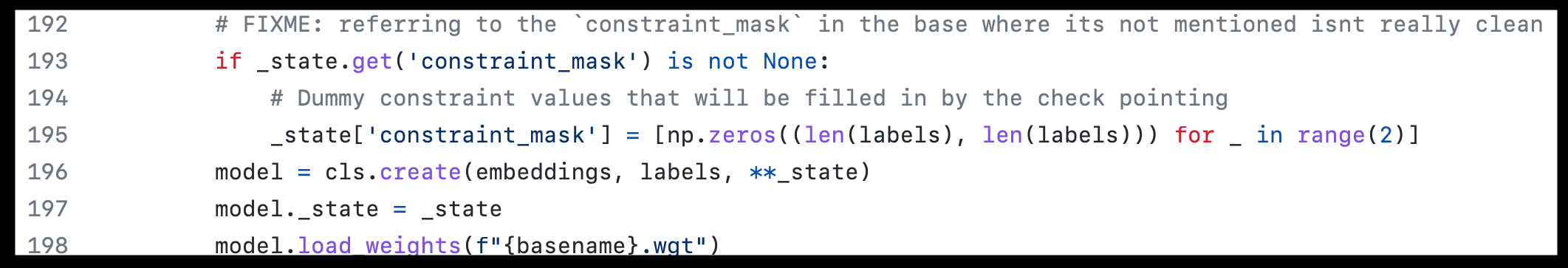}
  \caption{ Example of SATD in a Natural Language Processing application, \textit{Mead-ML}.
  }
  \label{fig.debt.eg}
\end{figure}

TD in general is hard to automatically detect from software engineering resources. However, in an important subset of cases, developers explicitly admit having produced a somewhat temporary implementation that possibly requires more work in the future. Such debt is referred to as Self-Admitted Technical Debt (SATD), and can be identified within comments that are inserted in the source code while developing~\cite{sierra2019survey}. For example, Figure~\ref{fig.debt.eg} provides an example of a natural language processing application, \textit{Mead-ML}\footnote{Accessed on: 2024-03-22, \url{https://github.com/mead-ml/mead-baseline/commit/d4d8fab88a7424785ef508d0c1bd24ac58cae4d8\#diff-c8db4ed3ecd32411a0c3bfecf8555d4ca0cf2181bfcce78b670b9cdc79e0b99d}}, where the code has a ``dirty'' implementation of saving a dummy matrix of zeros. In lines \#190 and \#192, the developers admit that this ad-hoc implementation (TD) will be fully developed after ``check pointing" is implemented in the future. Another example of SATD declaration is the comment
``\textit{\# TODO handle attention history}"\footnote{Accessed on: 2024-03-22, \url{https://github.com/ufal/neuralmonkey/blob/master/neuralmonkey/attention/transformer_cross_layer.py}} within a code base that builds an attention model in the Neural Monkey project, since it indicates the missing implementation of handling model weights.

The problem of TD in ML is exacerbated due to the “implicit” requirements of ML software products, which are grounded in data rather than specification documentations. As pointed out by Sculley et al.~\cite{Sculley}, the problem of TD is pronounced in ML due to the stochastic nature of ML, as opposed to the deterministic nature of traditional (non-ML) software, leading to quality issues similar to traditional, non-ML software as well as new types of issues. For instance, an SATD in an ML model’s hyper-parameter configurations or data pre-processing will result in an inferior model with lower accuracy. Every incorrect decision made by such a model may potentially negatively affect various aspects such as the vendor’s reputation, user experience, or financial performance. 
Building on the insights of Sculley et al.~\cite{Sculley}, this paper sets out to empirically validate whether the technical debt manifestations they describe in earlier work are indeed prevalent in real-world ML applications-and if so what are their characteristics. Empirically understanding the nature of SATD via code comments as recommended by Potdar and Shihbab allows us to explore direct insights into the acknowledged shortcomings in the implementation of ML systems. Since the nature of code changes performed in ML pipelines is different than that of the traditional non-ML software~\cite{bhatia2023towards}, it is only reasonable to expect that SATD in ML code should appear distinct. Hence, we use code comments as a starting point, whereas other aspects of ML technical debt, such as debt in the underlying data, or model artifacts (perhaps embedded in project documentation, architectural design documentation, or project management systems) fall outside the scope of this current study. This delineation lays the groundwork for future research to explore such avenues.

Despite recent efforts to understand SATD introduction and removal in deep learning frameworks, we still do not know much about the diffusion and evolution of technical debt in ML-based systems using such frameworks. Neither do we know the composition or the circumstances of SATD introduction or removal. While several researchers have investigated the detection, evolution, and impact of technical debt in traditional software systems (e.g.,~\cite{Alfayez2020,Lim2012,liu2021exploratory,brown2010managing,wehaibi2016examining,letouzey2012managing,kruchten2012technical,kruchten2013technical,bavota2016large}), software systems powered by ML models (i.e., ML-based systems) have mostly remained out of scope.

As the contributions of this research, we present the first extensive empirical study to compare the prevalence of SATD between Machine Learning (ML) software and traditional software, analyzing 318 open-source ML projects to provide a comprehensive overview of SATD occurrence. Furthermore, our research extends the current taxonomy of SATD by identifying new categories specific to ML software development, including Configuration Debt and Inadequate Testing Debts, thus broadening the understanding of TD within ML software systems. Next, our study offers an analysis of SATDs mapped to various stages of Amershi et al.’s ML pipeline (such as data preprocessing, model training, and deployment), highlighting the phases most vulnerable to TD. Through survival analysis, we reveal the patterns of SATD introduction and resolution in ML projects, showing that SATDs are introduced and removed more swiftly in ML software compared to conventional software. Finally, we investigate the factors leading to the persistence of SATDs in ML software, providing insights into the attributes (e.g., complexity, change history) associated with enduring debts.

\section{Related work}\label{sec.relatedwork}
In this section, we present the literature related to our work.
\subsection{Studies on Software Engineering for Machine Learning}
Software Engineering for Machine Learning (SE4ML) is a recent and rapidly evolving field with research that encompasses of a broad set of topics like requirements engineering, ML design and architecture, testing and validation, explainability and interpretability, ethics and fairness, data management, model deployment maintenance and life-cycle management, training performance optimization, and ML security. \\
A systematic literature review by Nascimento et al.~\cite{nascimento2020software} highlighted the unique challenges in ML testing, ML software quality, and data management in ML systems. Moin et al.~\cite{moin2021data} proposed a novel approach to integrate Model-Driven Software Engineering and Model-Driven ML Engineering, with a focus on automated ML, aiming to assist software engineers without deep ML knowledge. Tantithamthavorn et al.~\cite{tantithamthavorn2018impact} emphasized the need for explainable ML in software engineering, making ML models more practical and actionable. Kästner and Kang~\cite{kastner2020teaching} discussed the importance of teaching software engineering skills to students with an ML background, focusing on realism, robust infrastructure, and ethical considerations. Singla et al.~\cite{singla2020story} analyzed the adoption of Agile methodology in machine learning projects, finding a higher number of exploratory tasks and difficulty in estimating task duration. \\
Within such a vast body of SE4ML research, our research expands on the ``deployment maintenance and life-cycle management'' area. This area is a ripe domain with prior research focusing on various aspects like code changes (e.g., Bhatia et al.~\cite{bhatia2023towards} examined the nature of code changes in open source ML software), ML asset management (Idowu et al.~\cite{idowu2022asset} focused on tracking, exploring, and retrieving ML assets), ML code smells (e.g., Jebnoun et al.~\cite{jebnoun2020scent} analyzed 10 code smells and their properties in ML software). In contrast, our study focuses on the SATD aspect and helps ML stakeholders understand the nature of TD and enable ML practitioners to gain awareness of TD in building ML applications.

\subsection{Empirical Studies on SATD in Traditional (Non-ML) Software}
    In 1994, Cunningham~\cite{cunningham1992wycash} introduced the concept of technical debt as ``quick and dirty'' work in software design and implementation resulting in ``not quite-right code''. However, TD is an abstract concept, as making tradeoffs between optimal software and meeting project deadlines is not quantitatively measurable. Hence, subsequently, Potdar and Shibab~\cite{potdar2014exploratory} extended the concept of TD to self-admitted TD (SATD), which was deemed to be a more visible measure of TD~\cite{nord2012search,kruchten2012technical,Alves2016}.
    Along with introducing the concept of SATD, Potdar and Shibab examined SATDs in large open-source projects - Eclipse, Chromium, Apache HTTP, and ArgoUML. They found a high prevalence (within up to 31\% of files) of SATD, and identified that more experienced developers add more SATD and are unable to pay off the technical development due to time pressure and code complexity.

    Building on the seminal work of Potdar and Shibab, multiple subsequent studies have identified SATD in traditional software. For instance, Maldonado et al.~\cite{maldonado2017empirical} studied the evolutionary aspects of traditional code and revealed that SATD lasts between 18 and 172 days in a system. Bavota and Russo~\cite{bavota2016large} performed a large-scale study of SATD across 159 software projects, contributing towards developing a taxonomy of SATD for traditional software. Wehaibi et al.~\cite{wehaibi2016examining} examined the impact of SATD on software quality, while Kashiwa et al.~\cite{kashiwa2022empirical} studied SATD in code reviews. 
    
    However, none of the prior studies investigate SATD in ML-based systems. As identified by the previous subsection, the engineering practices and research-based findings of traditional software cannot be expected to hold as-is for ML-based systems \cite{washizaki2019studying,ozkaya2020really,martinez2021software}, and hence, our research is important to identify the characteristics of SATD in the currently prevalent ML-era of software development. 
    
    Two studies bear similarities to our research. Firstly, akin to our study, the  study by Liu et al.~\cite{liu2018satd} explores SATD in deep learning frameworks by investigating the introduction and removal of SATD in seven deep learning frameworks. However, contrary to the study of frameworks, our research identifies debt in ML systems, which use the former frameworks to build and integrate pipelines into ML applications. Our study provides insights for a broader spectrum of ML stakeholders who do not deal with ML mathematical or algorithmic implementations, but need to manage ML assets (like data, model) for development of ML applications. 
    Secondly, the study by O'Brien et al.~\cite{obrien202223} mirrors our work closely. The authors check the different types of SATDs in ML software and check their distribution in different components of ML pipeline stages. While O'Brien et al.'s research is qualitative, we focus on both quantitative and qualitative aspects of SATD on a different dataset. Within the quantitative aspects, our work takes a genealogical approach to SATD, tracing the lifecycle from introduction to removal. Additionally, we put forth an exploratory model to understand the dynamics of long-lasting TD. Given our different dataset, our qualitative results complement the results of O'Brien et al. and provide a broader understanding of the prevalence and impact of debt across different datasets and domains.

\subsection{Studies Presenting SATD Detection Tools}\label{subsec.tools}
    As mentioned in our previous subsection, the domain of SATD detection is a fertile field for research, and has various tools proposed to tackle this challenge. 
    
    A few notable examples are the efforts made by Malando et al.~\cite{da2017using}, who constructed an NLP-based SATD classification model and tested their tool across ten substantial open-source projects. Later, in 2017, Huang et al.~\cite{huang2018identifying} expanded the scope by developing a predictive SATD detection ML model that utilizes feature selection to provide the important words from the comment text used for SATD identification. 

    As an outcome of this research, the authors presented a composite SATD identification tool that combines multiple classifiers from different source code projects. Subsequently, Ren et al.~\cite{ren2019neural} utilized neural networks for SATD identification by leveraging the Convolutional Neural Network architecture to deliver state-of-the-art performance in detecting SATD comments. Later, in 2020, Faris et al.~\cite{de2020identifying} enhanced SATD classification models by using a contextualized vocabulary that considers the level of importance of different textual patterns along with the relationship between patterns and debt types.

    Despite such numerous efforts in SATD detection, the practical adoption of these tools presents challenges. Many of these detection tools, despite their academic promise, exist only as prototypes and may lack maintenance or contain bugs that prevent their practical use. 
    
    We employed the text mining-based SATD detector tool developed by Liu et al.~\cite{liu2018satd} in our study, which leverages natural language processing and machine learning to classify SATD comments. We selected Liu et al.'s SATD detector tool because of its frequent maintenance and demonstrated applicability in other authoritative SATD studies (e.g., ~\cite{liu2021exploratory} and \cite{Zampetti}).

\section{Research Questions}\label{sec.rq_motivation}
In this section, we discuss the research of our study and along with the motivation to study them in the context of our Introduction. 

\subsection{\rqone}
    Sculley et al.'s seminal work \cite{Sculley} highlighted the tendency of ML programs to incur technical debt, remarking that the strong dependence of ML models on data makes it difficult to enforce strict abstraction boundaries. When building ML-based systems, developers often experiment with different configurations in search of the optimal model; building, testing, and comparing the performance of different prototypes to identify the best configuration leading to the most efficient model~\cite{amershi2019software}. This process can often result in multiple \emph{dead experimental code paths}, increasing the cyclomatic complexity of the code and the risk of an unexpected behavior arising from some of the obsolete experimental code paths~\cite{Sculley}. Another potential area where technical debt can accumulate quickly in ML code is in the model building or data processing configurations. Developers often have to handle a wide range of configuration options. For example, they often have to select among a variety of algorithm-specific learning settings, and--or from a large pool of pre-processing, post-processing, and verification techniques. In this research question, we aim to examine the extent to which ML software deployed today, contains technical debt, by comparing the prevalence of SATD in ML and non-ML code.
    While Sculley et al.~\cite{Sculley} elucidated additional ML challenges leading to accumulation of technical debt that are not present in non-ML software, empirical evaluations of said challenges still remain scarce. Hence, in this research question, we aim to examine the extent to which ML software deployed today, contains technical debt, by comparing the prevalence of SATD in ML and non-ML code.

    \subsection{\rqtwo}
    Bavota and Russo~\cite{bavota2016large} have developed a taxonomy of SATDs in traditional software systems. Given the specific nature of ML code as described by Sculley et al.~\cite{Sculley}, new types of technical debt not mentioned in this taxonomy may exist. Therefore, in this research question, we investigate the composition of SATD in ML-based software, by qualitatively investigating a sample of SATD comments in ML software.
    Our goal is to uncover types of SATDs that may be specific to ML code and identify the types of SATDs that are prevalent in ML code. Such an understanding can ML practitioners help be vigilant about the types of debt that exist in ML software.
    
\subsection{\rqthree}
    A typical ML software pipeline~\cite{amershi2019software} involves stages for reading data and cleaning data: 1) transforming data into features that can be consumed by the ML model; 2) training the model; 3) evaluating and validating the model; 4) and finally deploying the model~\cite{amershi2019software}. During the evolution of the system, engineering teams often have to update these stages to experiment with new models or adjust to changes in operation data distributions. Through this process, as observed by Sculley et al.~\cite{Sculley}, different types of technical debt can be introduced.
    In this research question, we examine the proportion of SATD contained in different stages of the studied projects by taking a sample of SATD comments.
    By knowing which stages are the most prone to technical debt, engineering teams will be able to better prioritize their maintenance activities and better support their ML engineers, e.g., with best practices and--or efficient tools. 

\subsection{\rqfour}

    Several empirical studies~\cite{Sculley,bavota2016large} report that technical debt leads to low quality, particularly in terms of maintainability, and makes further changes more expensive in the long run~\cite{potdar2014exploratory,nord2012search,kruchten2012technical,Alves2016}. Technical debt needs servicing, which may involve refactoring code, improving tests, and deleting dead code to enable future improvements and reduce errors. An understanding of the empirical estimates for such servicing of technical debt (TD) would help ML stakeholders better estimate project effort and timelines. For instance, the time taken for TD introduction indicates the fast-paced accumulation of issues like untested ML configurations, data mismanagement, or lack of robust ML practices. Conversely, the time taken to remove debt can be dependent on factors such as triaging debt based on its perceived urgency, developer availability, complexity in understanding and managing the ML process, or even breaking changes across other ML modules.
    Hence, by examining the timing of introduction and removal of SATD through survival analysis, and comparing these patterns between ML and non-ML projects, we aim to provide insights into how the unique characteristics of ML development impact the evolution of technical debt. \\
    Our analysis can aid in developing strategies tailored to ML for identification, management, and timely resolution of TD. We wish to provide development teams with a holistic view of SATD survival in ML software systems, allowing them to better prioritize maintenance activities and allocate their limited resources efficiently.

\subsection{\rqfive}
    Persistent long-term accumulation of technical debt often signals areas in the codebase suffering from insufficient maintenance~\cite{tom2013exploration}. Debt lingering in the codebase for an extended period is crucial to understand since it could be overlooked or ignored by developers based on their perceived lack of its importance, or perhaps the project maintainers continually defer addressing this debt to subsequent maintenance cycles. To avoid such practices, it is essential to understand the factors leading to this long-term accumulation of debt. This understanding can help ML development teams acquire the necessary information to redirect their focus towards debt more likely to linger for a long time, thereby preventing its long-term accumulation.    

\section{Case Study Setup} \label{sec.data}

\begin{figure*}[!h]
  \centering
  \includegraphics[width=\columnwidth, keepaspectratio]{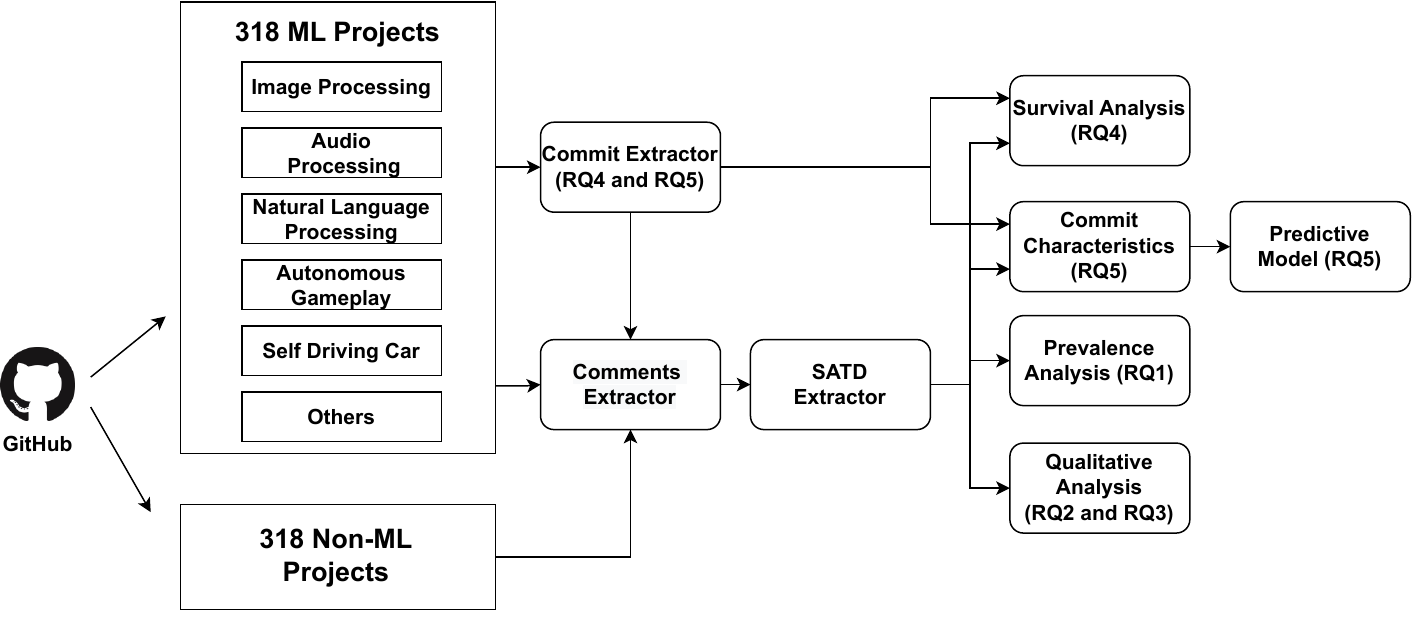}
  \caption{Data collection and processing steps.} 
  \label{fig.datacollection}
\end{figure*}

This section describes the design of our empirical study, which is also illustrated in Figure~\ref{fig.datacollection}. In the following, we elaborate on each of these steps in detail.

\subsection{Selection of ML Projects}\label{subsub.ml_procurement}

\subsubsection{Choice of ML Domains}
Applications of ML algorithms (ML-based systems) are prevalent in Healthcare, Finance, Retail, Travel, Social Media, etc.~\cite{dargan2020survey,deng2016deep,kamath2019deep}. All such industrial applications can be grouped into three high-level domains, namely, Image Processing, Natural Language Processing, and Audio Processing~\cite{das2015applications}. Recently, the use of reinforcement learning in Autonomous Gameplay has been gaining a lot of attention from academia and the industry~\cite{albrecht2018autonomous}, prompting us to include it in our dataset as well. Additionally, we include the Self Driving Car domain due to its widespread use in industry and the significant attention it has received from researchers~\cite{badue2021self}. The Self Driving Car domain mostly leverages computer vision in their perception modules and is an extended version of Image Processing~\cite{ma2020artificial}. While it is impossible to include all industrial domains, we select these five as representative examples~\cite{pouyanfar2018survey} of successful ML implementations. Finally, to account for other domains, we also add raw ML implementations like maintained ML courses that we could not fit into the five selected domains, labeling them as ``Other ML'' (see Figure~\ref{fig.datacollection}). Such repositories are ML implementations with a lot of community activity. For instance, the project ML-Glossary\footnote{Accessed on: 2025-04-23 \url{https://github.com/bfortuner/ml-glossary}} has 3.1+K stars, 700+ forks, 135 watchers, and the Readme reports ~54K monthly users---demonstrating strong community adoption. Understanding technical debt in machine learning courses is important; many practitioners initiate their ML journey through such courses, and the non-trivial adoption of these projects clearly represents their role in the ML open source ecosystem. It is possible that within these courses, comprehensive practices for managing technical debt might not be thoroughly covered. Further, by incorporating the management of technical debt from an early stage, learners can develop robust foundations, promoting best practices applicable in diverse, real-world scenarios.

\begin{table}[!h]
    \centering
    \caption{Queries used to find projects from various ML domains and a non‐ML domain, with rationale for synonym selection.}
    \label{tab.prj_queries}
    \begin{tabular}{p{3cm}p{4cm}p{6.5cm}}
    \toprule
    \textbf{Domain} & \textbf{Queries Used} & \textbf{Rationale} \\
    \midrule
    Self Driving Car 
    & ``autonomous driving machine learning'', ``self driving machine learning'' 
    & Reflects industry term (``autonomous driving'') vs common parlance (``self driving'') to capture variations in project naming. \\
    Audio Processing 
    & ``sound machine learning'', ``audio machine learning'', ``voice machine learning'', ``speech machine learning'' 
    & Covers raw acoustics (sound), general audio tasks, human speaker‐specific projects (voice and speech), ensuring broad coverage across subfields. \\
    Autonomous Gameplay 
    & ``autonomous gameplay machine learning'', ``game machine learning'', ``game reinforcement learning'',  
    & Captures varied terminologies (e.g. RL or ML) used by github projects for implementing ML in games\\
    Image Processing 
    & ``image processing machine learning'', ``vision machine learning'', ``image processing deep learning'' 
    & Captures classical terminology (image processing), domain context (vision), and DL‐centric methods for modern implementations. \\
    Natural Language Processing 
    & ``natural language processing machine learning'', ``chatbot machine learning'' 
    & Includes explicit NLP label and application‐specific term (chatbot) to catch transformer‐based conversational projects that may omit ``NLP'' tag. \\
    Other ML 
    & ``machine learning'' 
    & Serves as a catch‐all to include any ML repositories not covered by domain‐specific keywords. These were mostly educational projects. \\
    \bottomrule
    \end{tabular}
\end{table}

        \subsubsection{Keyword-Based Searching to get Candidate Repositories}
        For our analysis of the above six domains, we gathered ML-based open source projects from GitHub\footnote{Accessed on: 2025-04-23, \url{https://github.com/}}. We used the GitHub Search API v3\footnote{Accessed on: 2025-04-23,\url{https://docs.github.com/en/free-pro-team@latest/rest}} to look for keywords specific to each domain (see Table~\ref{tab.prj_queries}) and obtain a list of relevant projects.  To ensure the relevance of a repository to ML, we appended each domain's keywords term to ``machine learning''. Given that the Github Search engine retrieval is based on keyword matching in the repository name, description or contents of the readme file\footnote{Accessed on: 2025-04-23, \url{https://docs.github.com/en/search-github/getting-started-with-searching-on-github/understanding-the-search-syntax}}, we used a comprehensive set of keywords for each domain to ensure a high recall (i.e., from all available repositories pertaining to a topic, we should not miss any) in our selection process. We also ensure high precision by employing a comprehensive filtering process explained in the subsequent subsections. While our selected keywords capture the generic and specific naming convention within each domain, detailed rationales for each domain`s keywords are Table~\ref{tab.prj_queries}. 

        As suggested by prior studies~\cite{kalliamvakou2014promises, mirhosseini2017can, coelho2020github}, for each set of keywords, we identify the top matching projects by using the ``\textit{best match}'' feature of GitHub search API. For each search query, we leveraged the GitHub search API to look for the top 50 repositories, then moved on to the other keywords for that domain.
        
        \subsubsection{Filtering of toy-projects}
        Based on previous studies~\cite{kalliamvakou2014promises}, we filtered out ``toy projects'' from our comprehensive list of relevant projects for each domain, using a two level approach. At the lowest level, we employed a threshold on typical project-related metrics to filter out ``toy-projects'', which is a fairly common practice data collection strategy in Software Engineering research. For example, to study feature toggles in python projects, Hoyos et al.~\cite{hoyos2021removal} used Github Search API v3 to obtain a list of relevant projects, then filtered them out based on programmatic and manual checks. Among many others, Dabic et al.~\cite{dabic2021sampling} used a threshold of 10 stars to qualify matured projects. Alfadel et al.~\cite{alfadel2021use} used top-most starred, non-forked projects with at least 20 commits for studying dependabot security pull requests, while Chen and Jiang~\cite{chen2020studying} selected non-forked projects having at least 30 stars for studying the rationale behind using logging practices in Java (e.g., internationalization of the log messages).

        On top of this threshold-based filtering, our second level filtering further refines our sampled repositories using manual checks of the selected repositories based on a systematic approach commonly used in empirical software research. For instance, to study the characteristics of AIOps projects, Aghili et al.~\cite{aghili2023studying} obtained their dataset by searching GitHub using the keyword ``AiOps" and filtered the resulting repositories through manual verification to determine whether each repository actually focused on AIOps, a process akin to our research.
        
        Below, we present the criteria (\textbf{\texttt{C.}}) used to filter relevant projects at the Level-1 (programmatic check) and Level-2 (manual check) along with the rationale (\textbf{\texttt{R.}}) to apply that criterion.

        \textbf{Level-1: Programmatic Check}
        \begin{itemize}
            \item \textbf{\texttt{C1.}} A project should not be a fork of another project.  \\
            \textbf{\texttt{R1.}} Forks are copies of a given repository with minor variations. To avoid duplication and maintain originality, removing forked projects is necessary, as done by other research~\cite{bhatia2023towards,dabic2021sampling}.

            \item \textbf{\texttt{C2.}} A project should have at least five python source code files.\\
            \textbf{\texttt{R2.}} This threshold ensures that the selected projects have a minimal level of complexity and substance. 

            The use case of the research determines the choice of metric on which a threshold is applied.
            For instance, Hoyos et al.~\cite{hoyos2021removal} employed thresholds on data size, requiring the selected projects to have a size of more than 1MB. Aghili et al.~\cite{aghili2023studying} used a threshold of $\#stars>1$, while Dabic et al.~\cite{dabic2021sampling} used $\#stars>10$. 
            
            For our case, as indicated by Amershi et al.~\cite{amershi2019software}, ML pipelines contain stages like data processing, model building, etc. Given our focus on mature ML projects, we heuristically determined that such projects should typically implement the different stages of Amershi's ML pipeline, and do so in separate modules. Therefore, a project must include at least five Python files corresponding to these stages to be considered mature for our study. 
            
            \item \textbf{\texttt{C3.}} A project should have more than one month of development history.  
            \textbf{\texttt{R3.}} A minimum history of one month ensures that the project has undergone sufficient development cycles, potentially including updates and bug fixes, which are indicative of active use and maintenance. 
            Our choice of one month filters out projects where developers might have committed code for experimentation with certain ideas or new technologies, without the intention of ongoing development or maintenance. Such experimentation is common in the ML-development paradigm, hence our choice to include this criterion. 
        \end{itemize}

\textbf{Level-2: Criteria for manual filtering}
        \begin{itemize}
            
            \item \textbf{\texttt{C4.}} A project should implement a machine learning pipeline. \\
            \textbf{\texttt{R4.}} Our carefully designed search keywords (see Table~\ref{tab.prj_queries}) including the keyword ``machine learning” should ideally enable the GitHub search API to retrieve ML‐related repositories. However, there may be noise in this process, i.e., a non‐ML repo might have been recommended by GitHub’s search engine. This is why criterion C4 is in place.

            \item \textbf{\texttt{C5.}} A project should not be a machine learning library. \\
            \textbf{\texttt{R5.}} We are interested in ML-based systems, and not the ML libraries that are used by ML-based systems to build ML models. Hence, we exclude such libraries from our list of repositories.
            
        \end{itemize}
The verification of Level-2 criteria \textbf{\texttt{C4}} and \textbf{\texttt{C5}} was conducted through manual analysis by the first author. Our structured process for Level-2 manual checks unfolds as follows:
\begin{enumerate}
    \item \textit{Top-level check:} Each project's README was reviewed. As identified by Prana et al.~\cite{prana2019categorizing}, READMEs include sections on the project's purpose, installation instructions, and usage examples, providing clear information about the project's intent.
    
    \item \textit{Intermediate check:} 
    Intermediate check: If information regarding \texttt{C4} and \texttt{C5} was not discernible from the previous step, we proceeded to examine the files in the directory for the presence of ML system stages. Heuristically, the implementation of an ML system involves filenames being indicative of Amershi’s ML pipeline stages (e.g., model.py, train.py, or preprocess.py)~\cite{amershi2019software}.

    \item \textit{Code check:} If the previous steps did not yield valuable information, we conducted a finer-grained examination, which entailed 1) checking for the presence of ML library imports (e.g., tensorflow, sklearn, keras) and 2) inspecting for ML implementation specifics. The rationale for checking ML imports is based on the fact that ML-based systems utilize these libraries to build ML pipelines.
\end{enumerate}

    \begin{figure}
      \centering
      \includegraphics[width=0.7\columnwidth, keepaspectratio]{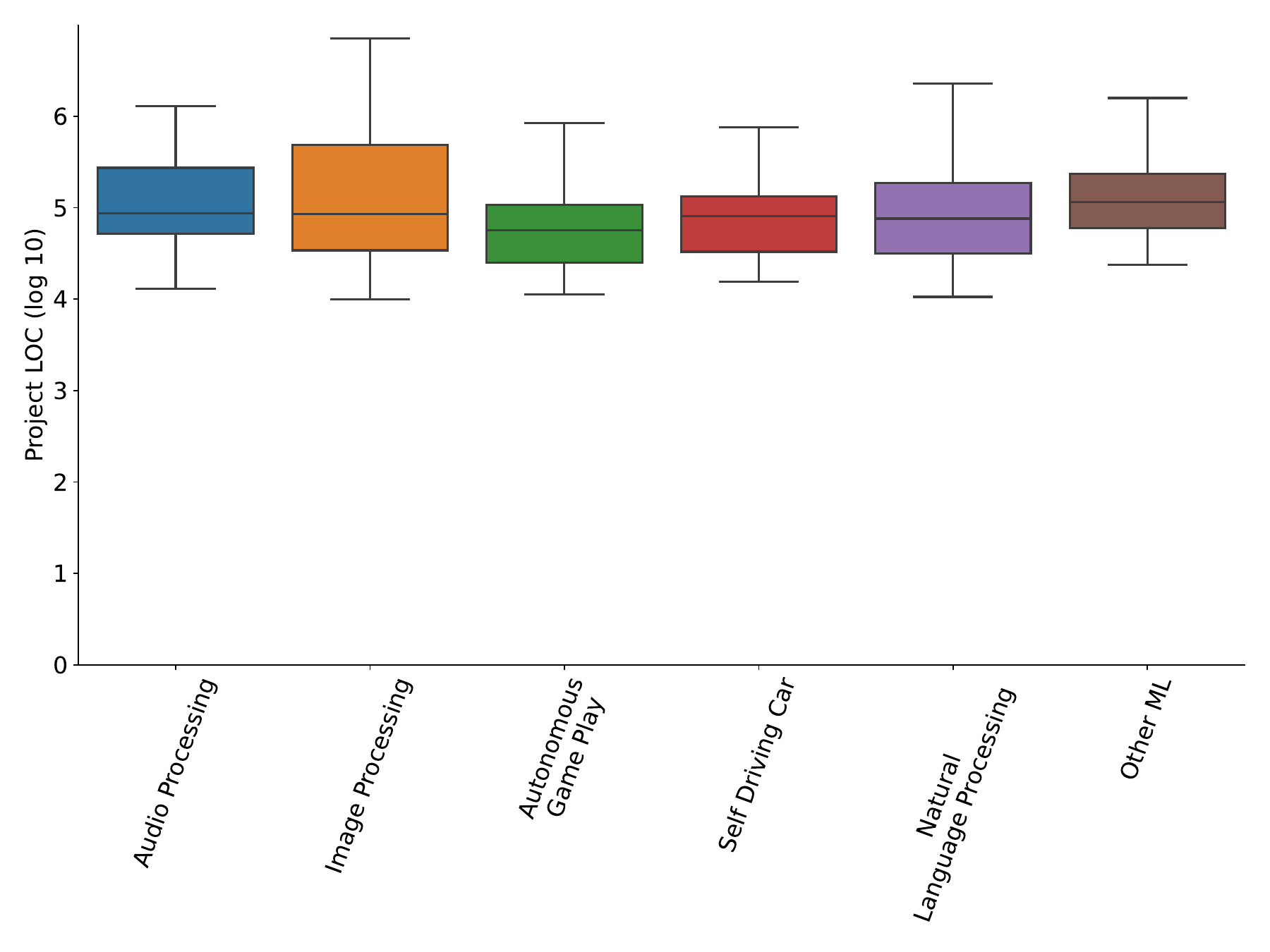}
      \caption{Distribution of number of \textit{Lines of code} for different project domains.} 
      \label{fig.loc-domain}
    \end{figure}
    
    Although this manual check was performed by one author, the nature of the check is objective (i.e., checking the presence of something); hence, we believe that the risk of subjective bias in an objective check is minimal.

    After applying the five selection criteria to our list of projects for each of the five domains, we obtained a total of 318 projects on which we perform our analysis. 
    
\subsection{Selection of Non-ML Projects}
    To answer the \textbf{RQ1} and \textbf{RQ4}, we also need a set of non-ML projects as a baseline. To allow a fair comparison, we employ a consistent selection process for both ML and non-ML projects.

    Specifically, we followed the sampling process outlined in Subsection~\ref{subsub.ml_procurement} and collected 318 non-ML Python projects from GitHub (matching the count of ML projects). The C1, C2, C3, and C5 ML criteria are directly reused for the non-ML set, whereas, the C4 criterion is adapted so that, while the ML criterion specifies that repositories \textit{must implement} ML, the non-ML criterion \textit{explicitly excludes} ML implementations.
    
    Using the same keyword-based selection mechanism of ML projects, we used the keywords ``python projects'', ``server'' and ``'' (empty keywords) to obtain a list of non-ML python projects sorted by the ``\textit{best match}'' criterion of GitHub Search. The rationale for using ``server'' is that this keyword often indicates backend implementations and is a non-specific term, i.e., the term ``server'' can have a diverse meaning in different software contexts, making it not specific to a technology, but a general term suitable for identifying non-ML repositories. Such projects often entail complex architectural decisions, configuration management challenges, and performance optimization issues, which are fertile grounds for the accumulation of technical debt.
    
    Similar to the filtering of the ML repositories (see Subsection~\ref{subsub.ml_procurement}), the three programmatic checking criteria (\texttt{\textbf{C1., C2.,}} and \texttt{\textbf{C3.}}) were validated, after which, the first author manually inspected the non-ML repositories to objectively verify the following checks:
     1) they do not contain ML components and 2) the non-ML repository does not implement a library. The three levels of checks, i.e., ``top-level'', ``intermediate-level'', and ``code check'' mentioned in the previous Subsection~\ref{subsub.ml_procurement} were done to fully validate the two non-ML criteria.

\subsection{Project Characteristics}
    Table~\ref{tab.data} provides a description of the selected projects along with their corresponding total number of python source code files per domain. Moreover, Figure~\ref{fig.loc-domain} plots the distribution of the projects' LOC for the different project domains. We find that the median project size is 80,477 lines of code across the entire ML project dataset, indicating that the selected projects are sizeable and matured.

\begin{table}[!h]
\centering
\caption{Distribution of the projects across the five categories.}
\label{tab.data}
\begin{tabular}{lrr}
\toprule
\textbf{ML Domain} & \textbf{Project Count} & \textbf{Python File Count} \\
\midrule
Self Driving Car & 39 & 1,701 \\
Audio Processing & 38 & 2,484 \\
Autonomous Game Play & 50 & 1,254 \\
Image Processing & 60 & 4,393 \\
Natural Language Processing & 78 & 3,357 \\
Other ML & 53 & 5,446 \\
\bottomrule
\end{tabular}
\end{table}

\begin{figure*}[!h]
    \centering
    \subfloat[Project Lifetimes]{{\includegraphics[width=6.5cm]{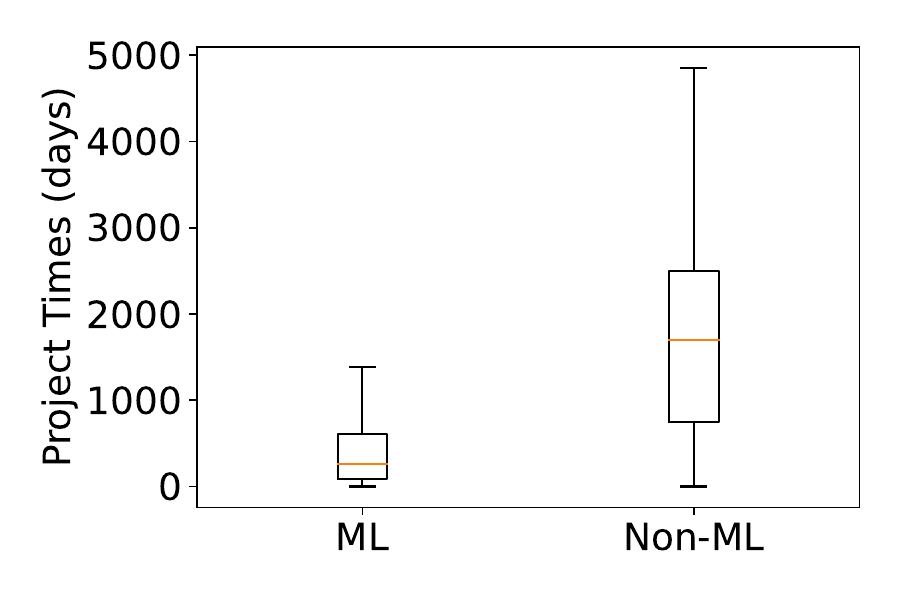} }}
    \qquad
    \subfloat[Contributor Count ]{{\includegraphics[width=6.5cm]{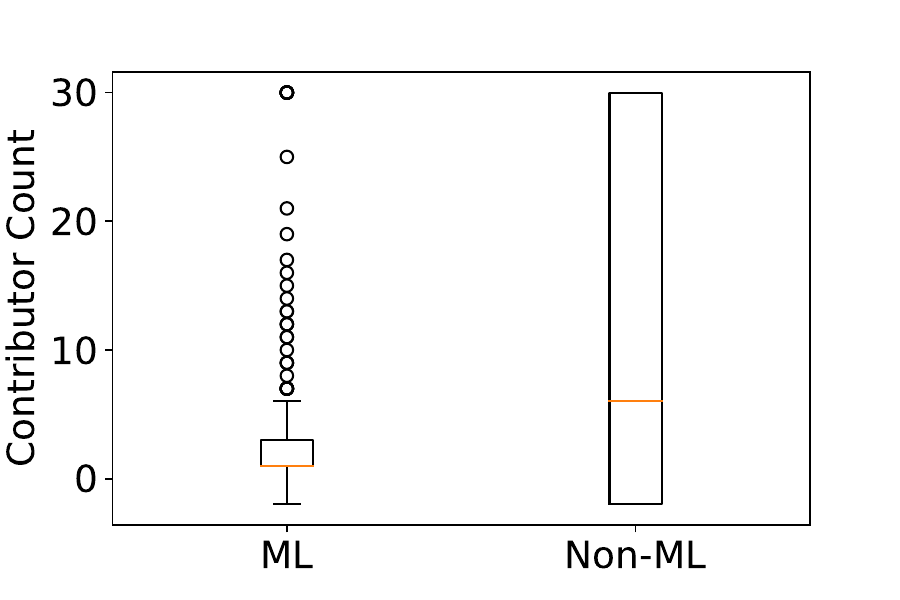}}}
    \\
    \subfloat[Commits count]{{\includegraphics[width=6.5cm]{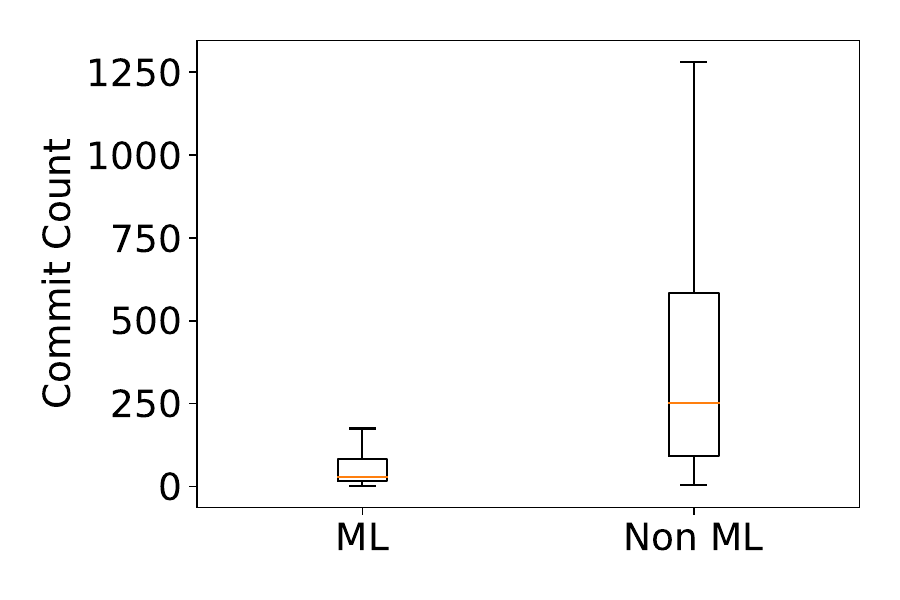} }}
    \qquad
    \subfloat[Churn per commit ]{{\includegraphics[width=6.5cm]{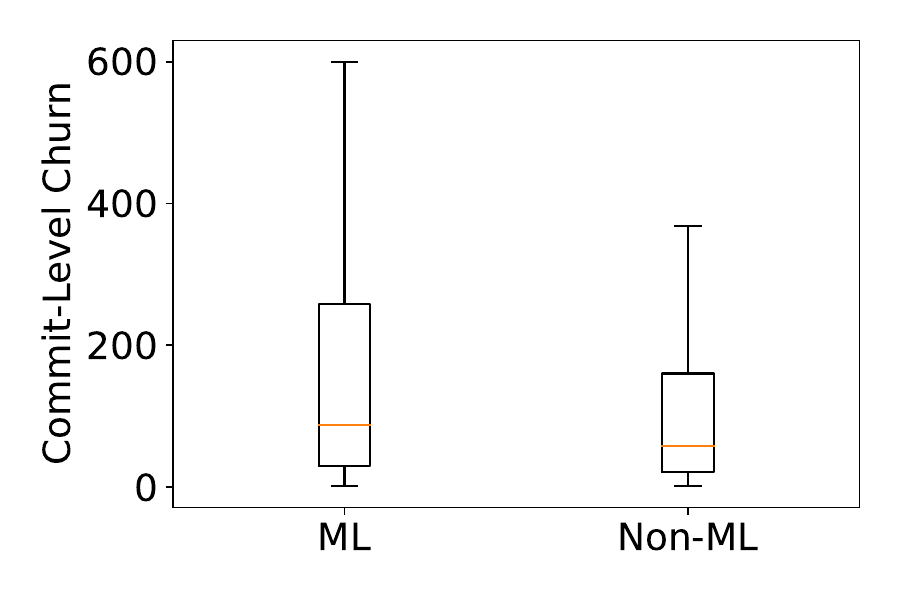} }}
    \caption{Project history lifetimes, commits, and churn per commit for ML and non-ML projects of our dataset.}
    \label{fig.ml_non_ml_comparison}
\end{figure*}

Figure~\ref{fig.ml_non_ml_comparison} compares the characteristics of the collected 318 ML and 318 non-ML projects. Since the boom in machine learning systems is fairly recent, ML projects naturally have shorter timelines (\ref{fig.ml_non_ml_comparison}-a) and less commits (\ref{fig.ml_non_ml_comparison}-c) as compared to non-ML projects. Similarly, ML projects have less contributors (\ref{fig.ml_non_ml_comparison}-b). Yet, surprisingly the total amount of developmental changes in ML seems to be much higher (absolute) churn compared to non-ML (\ref{fig.ml_non_ml_comparison}-d). 
Although the maturity level of ML projects is lower (since the ML era recently burgeoned), the fairly large amount of code modifications in each commit (churn) indicates that the ML project have high development traction, and hence a sizeable likelihood of introducing SATD. All comparisons in Figure~\ref{fig.ml_non_ml_comparison} are statistically significant (Wilcoxon Rank-sums test with p-value\textless0.01), with large effect sizes (Cohen's $\delta$\textgreater0.7). 

\subsection{SATD Detector Tool}
As stated in our literature review Section~\ref{subsec.tools}, many SATD detector tools used to identify instances of SATD from code comments. For our study, we use the state of the art SATD detection tool\footnote{Accessed on: 2024-03-22, \url{https://github.com/Tbabm/SATDDetector-Core}}, proposed by Liu et al.~\cite{liu2018satd}, which  employs a NLP-based pre-trained text classification ML model to automatically detect the presence of SATD in a code comment. As compared to other contemporary SATD detection tools that rely on pattern matching, this tool demonstrates a higher recall~\cite{liu2018satd}. 

This tool's use of an NLP-based model to determine whether a specific comment signifies SATD contrasts most of the other approaches, which merely scan for words like ``todo'' or ``fixme'' using string matching techniques. Therefore, the tool can successfully identify regular comments such as ``300 iterations seems good enough but you can certainly train longer'' or ``naive approach for generating forward mapping this is naive and probably not robust'' as SATD even when such strings lack an explicitly defined ``TODO''. This SATD detector has been utilized in  previous studies. For instance, Liu et al.~\cite{liu2021exploratory} used this tool for studying SATD in deep learning frameworks, while Zampetti et al.~\cite{Zampetti} used this tool for recommending when design technical debt should be self admitted. 

\subsection{Case Study Approach}

    \begin{table}[h]
    \centering
    \caption{Distribution of the dataset for machine learning and non-machine learning software.}
    \begin{tabular}{lrr}
    \toprule
                            & \multicolumn{1}{c}{ML Software} & \multicolumn{1}{c}{Non-ML Software} \\
    \midrule
    \#Projects              & 318                            & 318                                 \\
    \#Python-Files          & 20,319                         & 65,748                              \\
    \#Comments              & 424,248                        & 1,505,458                           \\
    \bottomrule
    \end{tabular}
    \label{tab.ML.nonML.dist}
    \end{table}
    
\label{subsec.approach.rq1}\subsubsection{\rqone} \leavevmode

    \noindent We extract 318 ML and 318 non-ML python projects following the description from Section~\ref{sec.data}.
    For each ML or non-ML project, we extract comments using \textit{Comment-Parser}, a PyPi framework\footnote{Accessed on: 2024-03-22, \url{https://pypi.org/project/comment-parser/}}. We obtained 0.4 Million and 1.5 Million comments in ML and non-ML software respectively. 
    A detailed description of the distribution of ML and non-ML software parameters is provided in Table~\ref{tab.ML.nonML.dist}. We qualify whether a comment is SATD using the SATD  detection tool\footnote{Accessed on: 2024-03-22, \url{https://github.com/Tbabm/SATDDetector-Core}}. 
    We perform sampling with replacement (bootstrapping) and compare the amount of SATD in ML and Non-ML projects across 1,000 runs. Since the number of comments in ML and Non-ML projects are disproportionate (see Table~\ref{tab.ML.nonML.dist}), bootstrapping prevents our results to be dependent on any one-specific sample seed, keeping the Standard Error ($\mathrm{SE} = \frac{s}{\sqrt{n}}$) of our setup as low as 0.07\% for ML comments and 0.05\% for non-ML comments. 
    
    We chose to perform our analysis on non-ML projects instead of reusing results previously published, to allow for a fair comparison. In fact, parameters like the programming language (Python vs Java in prior studies), the SATD detection tool, and the number of projects can impact the comparison.
    Nevertheless, we will refer to prior findings whenever relevant.

\paragraph{Approach to identify the rationale for high debt prevalence in ML vs. Non-ML: Advocatus Diaboli.}
A common approach to challenge prevailing assumptions and explore alternative explanations for an empirical observation is the Advocatus Diaboli (a.k.a. the devil’s advocate) approach~\cite{vysali2020quantifying}.  This approach is a recognized methodology used in prior software engineering and broader scientific research to systematically test alternative hypotheses. For instance, Minelli et al.~\cite{minelli2015know} used it to question whether their findings generalized beyond a Smalltalk IDE; Ferré and Rudolph~\cite{ferre2012advocatus} applied it to falsify the completeness of ontologies via logically absurd counterexamples. In line with these works, we apply this technique not to claim causation via metrics like churn or complexity, but to challenge the robustness of our own explanation by actively testing plausible alternatives.

In our case, we adopt the devil’s advocate perspective to examine the high debt prevalence in ML and Non-ML settings by assessing three potential predictors: 1) churn, 2) number of file changes, and 3) complexity.  While other metrics could be relevant, our selection is designed to provide an evaluation of these particular variables, ensuring a focused and manageable scope for detailed investigation. Below, we present the rationale to  select our metrics in the context of technical debt. 

\begin{itemize}
    \item \textit{Churn.} High churn indicates frequent changes and adjustments, which we hypothesize could potentially lead to a breeding ground for technical debt. High churn is associated with other poor programming practices and design issues, including higher likelihood of defects and smells~\cite{amit2020corrective,mauerer2021search,Morales15}. Employing the broken window theory~\cite{leven2024broken}, high churn signifying other aspects of poor design could also be extended to technical debt.

    \item \textit{Number of File Changes.}
    The count of changed files indicates the diffusion of changes. For instance, if multiple files are updated, it suggests significant changes, possibly indicating a breaking change or extensive refactoring activities. We hypothesize that such substantial changes could be fertile grounds for the accumulation of technical debt. 

    \item \textit{Complexity of File.}
    Complexity directly hampers the understandability and thus the maintainability of code. Research has indicated that quality attributes like debugging effort, security, vulnerability, refactoring effort, fault density, fault proneness, and maintainability are impacted by complexity~\cite{nguyen2017impact}. Hence, high complexity code is hypothesized to be prone to accruing technical debt. 

\end{itemize}
Normalizing churn by file size allows for a better understanding of whether changes are indeed “massive”. For instance, a high churn rate may be misleading if the file size is also large, as a file with more lines of code can naturally expect more lines to be updated, even during regular maintenance activities. Therefore, dividing the churn by the file size, i.e., total lines of code (LOC), provides a more accurate indicator of churn. We apply the same normalization denominator (i.e., LOC) to the other two metrics as well for a fair comparison.

\subsubsection{\rqtwo} \leavevmode

    \noindent We apply stratified sampling across our five domains using a 99\% confidence level and 5\% confidence interval and randomly sampled a total of 611 SATD comments from the SATD data set obtained in \textbf{RQ1}. We apply stratified sampling to avoid biasing our dataset towards SATDs from a specific domain. 
    Prior to classifying the sampled 611 SATD instances, the first two authors together examined an initial set of 50 samples (separate from the 611 samples) to build a common understanding of the SATD categories. 
    
    They used card sorting to categorize the SATDs using the extended taxonomy proposed by Bavota and Russo~\cite{bavota2016large}. The latter taxonomy includes SATD types by Bavota along with the addition of 1) ``On-Hold Debt"~\cite{Rungroj}, and 2) ``Refactoring'' under ``Code Debt''~\cite{suryanarayana2014refactoring} (see Figure~\ref{fig.debt_comprehensive_taxonomy}).
    
    In cases where the SATDs could not be categorized using the labels proposed by this taxonomy, a new category was created by us. A similar process was adopted by Bhatia et al.~\cite{bhatia2023towards} for the expansion of a prior taxonomy of code changes by Hindle et al.~\cite{hindle2008large} to involve the ML types of code changes. Akin to our research, Bhatia et al.~\cite{bhatia2023towards} employed card sorting, introducing new types of changes as additional codes whenever the existing labels did not adequately describe the changes.
    
    After this initial step, the authors separately classified another set of 65 samples (from a pool of 611), then computed the inter-rater agreement on this new sample using Cohen Kappa, and obtained a value of 94\%, which is very high. After achieving this high level of agreement, the first author manually classified the remaining 546 SATD comments. A similar approach was used by Kononenko et al. \cite{7886977} to classify code review survey responses, and Uddin and Khomh \cite{8643972} to classify API aspects.
    
    During the classification process, we encountered SATD comments containing the keyword ``\texttt{\#TODO}'' without any explanation of the task to be done. We labeled these instances as \textit{Undefined}. We also found instances for which we were unable to identify the exact type of technical debt reported. We labeled such instances as \textit{Unknown}. Overall,  40\% of SATD comments were considered to be either false positive or to belong to the categories \textit{Undefined} or \textit{Unknown}. To answer our research question, we remove these cases and calculate the percentage distribution of different types of SATDs from the remaining dataset of 367 samples.

    \subsubsection{\rqthree} \leavevmode

    \noindent We utilized the same dataset of 611 SATD comments generated for \textbf{RQ2}, with the objective of mapping these comments to specific stages of the ML workflow. Following the categorization established by Amershi et al.~\cite{amershi2019software}, we leverage five critical stages within the ML lifecycle, namely: \textit{data reading}, \textit{data preprocessing}, \textit{model building}, \textit{model validation}, and \textit{model deployment}. 
    A similar approach was followed by Foalem et al.~\cite{foalem2024studying} for mapping log statements to a corresponding ML pipeline stage.

    Since our goal is to map each of the 611 SATD comments to the preexisting ML stages (without modifying any label), we use a closed, single-label classification coding strategy to assign one ML stage (label) to a specific SATD instance. We also added a label of “Non-ML” indicating that a SATD instance was found in a non-ML part of the system. 

Specifically, to evaluate the ML stage at which SATD was present, the coders inspected the following data:
\begin{itemize}
\item \textbf{Filenames}: For instance, a SATD in a module called ``model.py'' is likely to occur in the ``model-building'' ML stage.
\item \textbf{Function blocks and calls}: The function/method name, along with the logic was evaluated to determine the ML stage. For instance, a function called ``\texttt{def get\_metrics():}'' likely indicates the model validation stage of the ML pipeline.
\item \textbf{Code logic}: Along with the clues from the filename and function name, we inspected the functional implementation of the code block where the SATD comment was made to obtain a functional understanding of the ML stage at which the SATD was present. 
\end{itemize}

If we were unable to find any information from the three aspects, we labeled the ML pipeline stage as “unknown”. This “unknown” category does not indicate the absence of debt, but instead the evaluators are unable to localize its exact ML stage. 

The use of the three aspects presented above is reflected in the following case: A sample SATD comment found in the unmanned autonomous project flywave\footnote{Accessed on: 2024-03-22, \url{https://github.com/UAVs-at-Berkeley/flywave}} at Line\# 516\footnote{Accessed on: 2024-03-22, \url{https://github.com/UAVs-at-Berkeley/flywave/tree/master//networking/bleConnection.py}} was present in the function called ``\texttt{send\_param\_command\_packet}''. Furthermore, the fact that this SATD was located in the file named ``\textit{networking\textbackslash bleConnection.py}'' points to this SATD instance pertaining to the ``Data Reading'' stage of the pipeline. The last heuristic, i.e., “inspection of code logic,” revealed the SATD's association with extracting telemetry data from drone sensors, further concretizing the ``Data Reading'' label.

To ensure reliability in our categorization, the first two authors conducted an initial pilot study on an initial set of 65 samples to independently provide labels, and later discussed and resolved any classification disputes until consensus was reached. The inter-rater agreement, measured by Cohen’s Kappa, was 81\% among the first two authors for a subset of 65 comments. This level of agreement, surpassing 70\%, is considered adequate~\cite{byrt1996good}. Following this pilot phase, we refined our categorizations by 1) eliminating any false positives, 2) removing the SATD instances labeled ``unknown'' and 3) removing the SATD from Non-ML components. Eventually, we found a total of 370 samples corresponding to different stages of the ML pipeline.

\subsubsection{\rqfour} \leavevmode

    \noindent We perform survival analysis~\cite{miller2011survival} to examine, for each project, (i) the time until the introduction of SATD and (ii) the time taken to remove SATDs after they are introduced in a file for the first time. The former analysis examines the survival of ML software projects concerning debt occurrence, while the latter analysis is about the survival of SATDs instances once they are introduced in an ML-based software project.
    
    Survival analysis \cite{miller2011survival} is a statistical technique for analyzing the expected duration of time until the occurrence of an event of interest or censoring of an event (if the event does not happen). The event of interest in our study is the occurrence of SATD across the entire project history, and the length of the observation window is the start of the project until the time at which we obtained the projects' latest snapshot at the time of performing the empirical study (Summer 2021). If the event is not observed during the observation period, the corresponding subject will be censored at the end of the period, which survival analysis is able to deal with.

    \begin{table}[!h]
    \centering
    \caption{Summary of events and corresponding variables for survival analysis.}\label{tab:survival.events}
    \begin{tabular}{@{}lcc@{}}
        \toprule
        & \textbf{SATD Introduction} & \textbf{SATD Removal} \\
        \midrule
        \textbf{Censoring} & File never had SATD & SATD never removed \\
        \textbf{Time to event} ($T$) & $\delta(T_{\text{SATD intro}} - T_{\text{File create}})$ & $\delta(T_{\text{SATD remove}} - T_{\text{SATD intro}})$ \\
        \textbf{Status} & Observed (1) / Censored (0) & Observed (1) / Censored (0) \\
        \bottomrule
    \end{tabular}
    \end{table}
    
    \textit{Time to event} and \textit{status} are two important variables for survival analysis. \textit{Time to event ($T$)} is the elapsed time between the beginning of the observation (start date of the project) and the occurrence of the event of interest or the censoring of data. In our study, this time is measured in \#days. Hence, $T$ is a random variable with positive values \cite{miller2011survival}. A summary of the Time ($T$), Event of interest, and Status is provided in Table~\ref{tab:survival.events}. For the introduction of SATD, $T$ is the elapsed time between the start date of the project to the time of contamination of the file by SATD. For the removal of SATD, $T$ is the elapsed time between the removal and the introduction of SATD. 
    
    \textit{Status} is a Boolean variable that indicates whether the event of interest is observed or whether the data is censored. 
    
    \textit{The Survival function S(t)} gives the probability ($P(T > t)$) that a subject will survive beyond time $t$. After arranging our data in increasing order of $T$, we plot the survival curve and estimate the survival probability using the non-parametric Kaplan-Meier estimator \cite{kaplan1958nonparametric}. The Kaplan-Meier estimation is computed following Equation \ref{eq:kapmai}, where $t_{i}$ is the time duration up to event-occurrence point $i$, $d_{i}$ is the number of event occurrences up to $t_{i}$, and $n_{i} $ is the number of subjects that survive just before $t_{i}$. $n_{i}$ and $d_{i}$ are obtained from the aforementioned ordered data. 
    \begin{equation}
    \label{eq:kapmai}
      S(t)= \prod_{i:t_{i} \leq t}{[1-\frac{d_{i}}{n_{i}}]}
    \end{equation}


\subsubsection{\rqfive}\leavevmode 

    \noindent In this RQ, we build a explanatory ML models to obtain important features used to classify long-lasting and quickly removed SATD. The choice of the explanatory metrics, mining those metrics, assigning the response variable, then building the ML model, validating the ML model, and finally interpreting the model are provided below.
    
    \paragraph{Selection of Metrics.}
    To understand the characteristics contributing to long-lasting debt, we capture metrics in the context in which SATDs are introduced. Specifically, we collect information about the code \emph{change} (i.e., LA\_{[}mod{]}, LD\_{[}mod{]}, CM\_{[}mod{]}, CT\_{[}mod{]}), the  diffusion of the change (i.e., LA\_{[}diff{]}, LD\_{[}diff{]}, LM\_{[}diff{]}, MD\_{[}diff{]}, MF\_{[}diff{]}), the complexity of the files involved in the change (Complexity, LOC, Tokens), and the timing of prior changes and SATD occurrences in the commit's files (i.e., Had\_SATD\_{[}hist{]}, PT\_{[}hist{]}, LCT\_{[}hist{]}, FC\_{[}hist{]}).
    These metrics are summarized in Table~\ref{tab.metrics}.
    
    Our dimensions `Change', `Diffusion', `Complexity', and `History', are widely employed in the domain of SDP (software defect prediction), a well researched field in Software Engineering. The rationale for their usage in the context of explaining SATD survival is based on the understanding that areas of code that undergo frequent changes, or are complex, are more prone to defects. Given that SATD indicates sub-optimal implementations that may not be outright defects but could result in compromised software quality, we consider using these metrics for SATD prediction. Moreover, prior research in SATD prediction by Yan et al.~\cite{yan2018automating} previously employed such dimensions (i.e., diffusion, history, and message) in their study. In contrast to Yan et al.'s research, we exclude the message dimension from our study as our focus is on assessing the attributes of persistent self-admitted technical debt (SATD) rather than SATD detection itself. 
    
     We explain the rationale for selecting each dimension below:
    \begin{itemize}
        \item \textit{Change.} As ML models are refined and improved, methods and functions may undergo frequent modifications. Rapid iterations and changes in ML codebases can introduce debt as developers make trade-offs between short-term expedient decisions for immediate gains.
    
        \item \textit{Diffusion.} As data and model architectures evolve over time, changes can propagate across multiple files and directories. A high count of changes and modifications in certain areas might indicate collaborative features or components undergoing significant evolution, signaling potential areas where SATD could arise due to integration or coherence challenges among contributions.
        
        \item \textit{History.}  `History' metrics like `PT\_{[}hist{]}` and `FC\_{[}hist{]}' lend insights into the temporal aspects of code changes. As ML projects evolve, certain modules can accumulate technical debt due to shifting project requirements or the rapid evolution of ML techniques. Monitoring the frequency and timing of changes in relation to SATD introductions can be valuable metrics for anticipating future debts.
    
        \item \textit{Complexity.} These metrics do not just capture control flow intricacies but also provide insights into the inherent complexity of ML-related tasks, such as feature extraction or model training routines. By analyzing the total token count or number of lines of code, we can gauge the intricacy of certain ML tasks to identify areas where SATD might be more prevalent due to the intricate nature of ML routines.
        
    \end{itemize}

\begin{table}[!h]
\centering
\caption{Metrics used to analyze the evolutionary patterns of SATDs and predict their occurrence.}\label{tab.metrics}
\begin{tabular}{>{\raggedright}p{1.6cm} >{\raggedright}p{2.5cm} p{8.4cm}}
\toprule
\textbf{Dimension} & \textbf{Name} & \textbf{Definition} \\
\midrule
\multirow{4}{1.6cm}{Change} & LA\_{[}mod{]} & \# Lines added during the modification. \\
& LD\_{[}mod{]} & \# Lines removed during modification. \\
& CM\_{[}mod{]} & Total \# ``changed methods'' during modification. \\
& CT\_{[}mod{]} & ``Change type'' of the modified file can be ``Add’’, ``Edit’’ or ``Delete’’. \\
\hline
\multirow{5}{1.6cm}{Diffusion} & LA\_{[}diff{]} & Total \# lines added for all modifications in the commit. \\
& LD\_{[}diff{]} & Total \# lines removed for all modifications in the commit. \\
& LM\_{[}diff{]} & Total \# ``lines modifications'' in the files in the commit. \\
& MD\_{[}diff{]} & \# ``Modified directories'' where source code files were modified. \\
& MF\_{[}diff{]} & Total count of Python files modified. \\
\hline
\multirow{3}{1.6cm}{Complexity} & Complexity & Cyclomatic complexity of the file after the modification. \\
& LOC & \# Total lines of code after the modification. \\
& Tokens & Total token \# after the modification. \\
\hline
\multirow{4}{1.6cm}{History} & Had\_SATD\_{[}hist{]} & Whether the modified file had SATD in its previous versions. \\
& PT\_{[}hist{]} & ``Project Time'' or time between the first and modification commit. \\
& LCT\_{[}hist{]} & ``Latest Commit Time'' or time between the latest and modification commit. \\
& FC\_{[}hist{]} & \# ``File Changes'' or number of prior commits where the file was edited. \\
\bottomrule
\end{tabular}
\end{table}

    \paragraph{Data Mining.}
    We use the Python framework PyDriller\footnote{Accessed on: 2024-03-22, \url{http://pydriller.readthedocs.io/}} to extract all commits submitted to the repositories of our studied projects. Each commit consists of a comment describing the reason for the commit and source code diffs describing the changes performed on source code files. We developed a custom parser to extract the changed comments accompanying the source code changes and used our SATD detector to identify technical debt admissions as described in Section \ref{subsec.rq1}. 
    We also use PyDriller to extract source code diffs and compute the metrics described in Table~\ref{tab.metrics}.
    
    \paragraph{Assigning Class Labels.}
    To identify long-lasting SATDs, we compute the time elapsed since the creation of the comment admitting the debt. We compute this elapsed time using information from the commit diffs (i.e., $++$ and $--$ lines of code comments for each source code file in the commit). Specifically, we calculate the time between the introduction of the SATD comment and its removal. From our collected data, we observed cases in which the same SATD commit is present multiple times within the same file, and removed these cases from our case study to avoid biasing our results with duplicated information. After computing SATD duration time for a total of 6,427 SATD instances in our dataset, we computed the quartiles of the obtained distribution and used the first and fourth quartile to identify \textit{``quick removal''} and \textit{``long lasting''} SATD respectively. In total, we obtained 2,595 SATD instances with duration times above 75\textsuperscript{th} percentile. We assigned them to the group of \textit{long-lasting} SATDs. We obtained the first quantile, i.e., 2,595 instances of with duration times below the 25\textsuperscript{th} percentile as the \textit{quick removal} SATD.
    
\begin{figure}[!h]
      \centering
      \includegraphics[width=0.7\columnwidth,keepaspectratio]{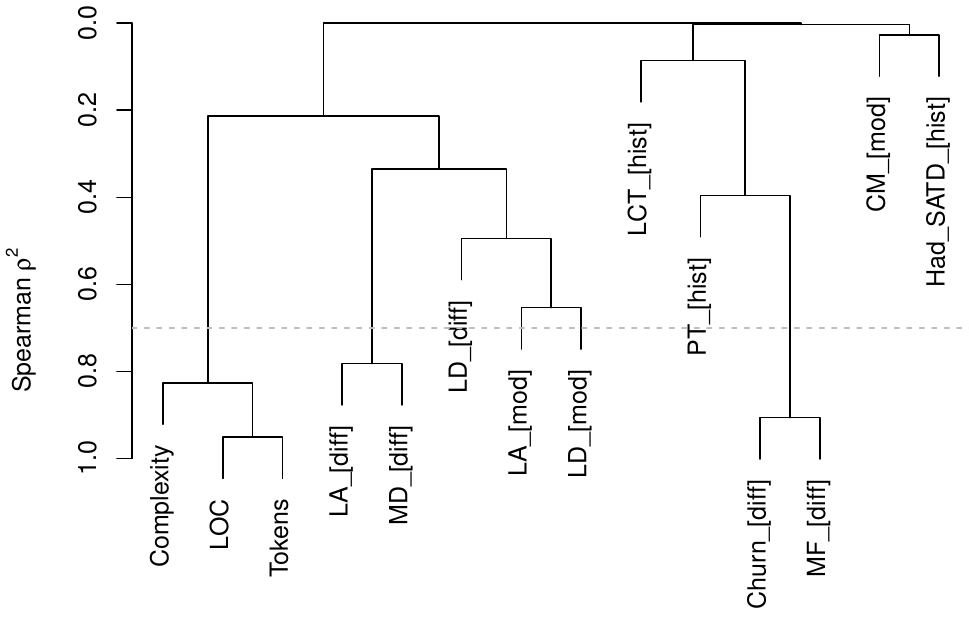}
      \caption{Correlation analysis dendrogram showing which pairs of features are correlated. Dotted line indicated a correlation threshold of 0.7
      }
      \label{fig.correlation}
    \end{figure}

    \paragraph{Correlation and Redundancy Analysis.}
    Before training our explanatory models, we performed correlation and redundancy analysis to prevent issues related to multicollinearity between variables. We used the R package Hmisc\footnote{Accessed on: 2024-03-22, \url{https://cran.r-project.org/web/packages/Hmisc/index.html}} to identify the correlated variables. We used the threshold of 0.7 on the Spearman's rank correlation between the exploratory variables. From an initial set of 14 metrics, we removed four metrics. A pictorial representation of our correlation analysis is provided in Figure~\ref{fig.correlation}. After removing the correlated metrics, LA\_[diff] was removed during the redundancy analysis\footnote{Accessed on: 2024-03-22, \url{https://www.rdocumentation.org/packages/Hmisc/versions/4.5-0/topics/redun}}. 
    
    Finally, the metrics: Complexity, Churn\_[diff], MD\_[diff] LD\_[diff], LA\_[mod], LD\_[mod], CM\_[mod], LCT\_[hist], PT\_[hist], and Had\_SATD[hist] survived our correlation and redundancy analysis.
    
    \paragraph{Model Building.}
    We chose the Random Forest classifier for our explanatory model. A Random Forest classifier is created using bagging of several decision trees and taking their aggregate value as the output of the classifier. We use Random Forest for its fast prediction speed via parallelizability, robustness against feature scaling, and robustness against outliers and non-linear data.
    We used the Python SciKit Learn\footnote{Accessed on: 2024-03-22, \url{https://scikit-learn.org/stable/modules/generated/sklearn.ensemble.RandomForestClassifier.html}} framework for building the models. 
    
    \paragraph{Model Validation.}
    To ensure the statistical robustness of our results, we perform a 100-out-of-sample bootstrapping. The approach involves sampling the training set from the whole dataset with replacement. The testing set comprises all the samples present in the original data but absent in the bootstrapped training data. We build and test models on each training and testing set to draw statistically robust conclusions as suggested by~\cite{tantithamthavorn2016empirical}

    \paragraph{Model Interpretation.}
    We use SHAP (SHapely Additive exPlanations), to interpret and explain our models. A Shapley value is the average marginal contribution of a feature value across all possible coalitions~\cite{shapley201617}. We compute these values using the Python framework, SHAP\footnote{Accessed on: 2024-03-22, \url{https://github.com/slundberg/shap}} which provides a collaborative game theory approach for the calculation of the average marginal contribution of each of the model's features.

\subsection{Replication Package}
All the scripts along with the mined data is provided in the replication package\footnote{Accessed on: 2024-03-22, \url{https://drive.google.com/drive/folders/1n-gwAxFANS-PPewnk1Qgmnwd0zMAr7Yb}
*This tentative g-drive link will be replaced by a github repository after the reviewing process. 
}.

\section{Case Study Results}\label{sec.results}
In  this  section  we  report  and  discuss  the  results  of  our research questions. For each research question, we present the  motivation, the approach, and discuss the results.  

\subsection{\rqone} \label{subsec.rq1}

\begin{figure}[!h]
  \centering
  \includegraphics[width=0.7\columnwidth,keepaspectratio]{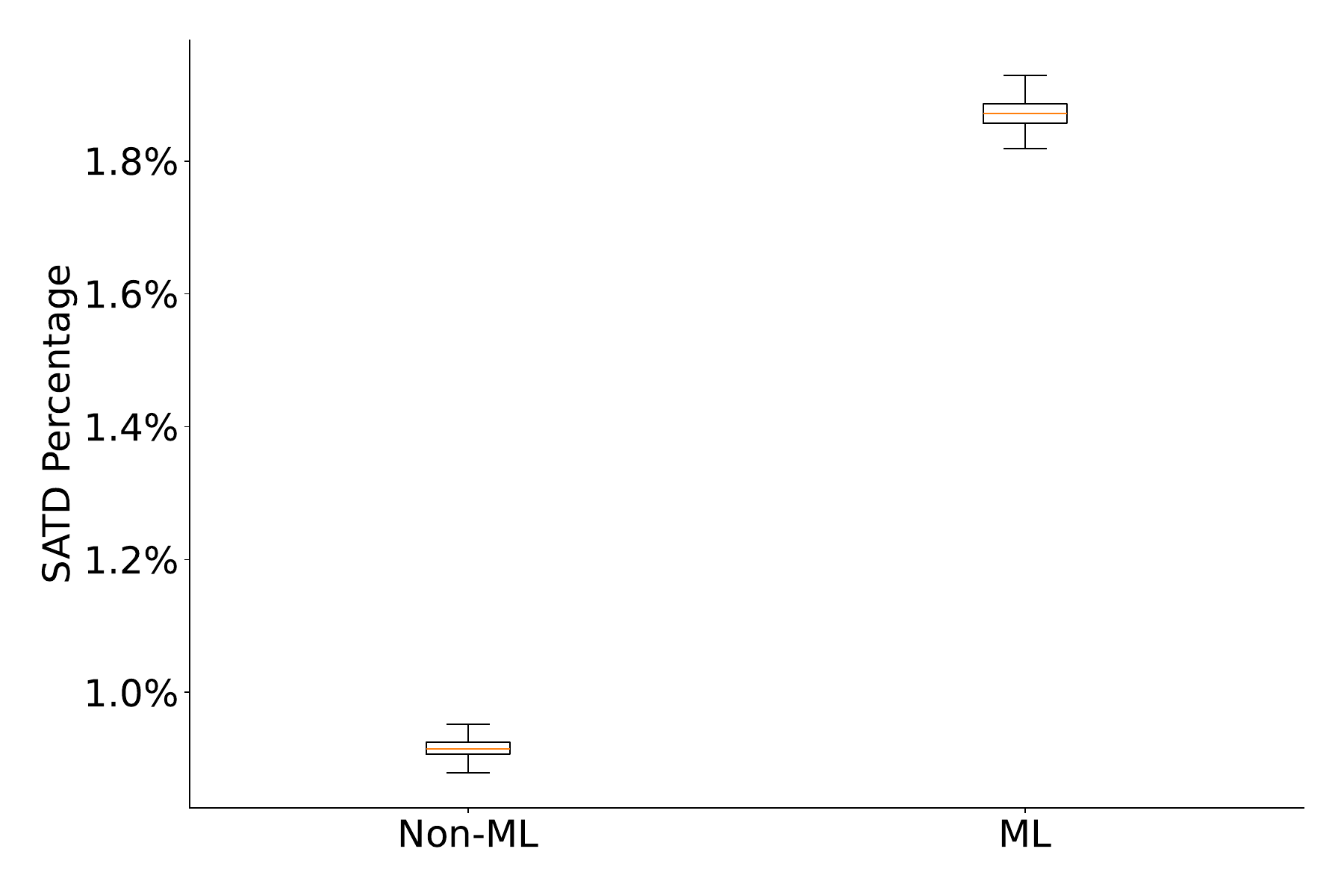}
  \caption{Percentage of SATD in ML and non-ML projects. The comparison has been made by collecting 424,248 bootstrapped samples from ML and non-ML comments and repeating the process 1,000 times (to ensure robustness).}
  \label{fig.debt.comparison}
\end{figure}

\noindent \textbf{Machine learning-based projects have 2 times more SATD in comparison to non machine learning projects.} Figure~\ref{fig.debt.comparison} shows the distribution of SATD in ML and non-ML software. The median of percentages of SATD in ML and non-ML software is 1.87\% and 0.92\%, calculated over 1000 times using bootstrapping, i.e., sampling with replacement of 424,248 (size of the ML-dataset) comments from both ML and non-ML categories. This indicates a 2 times more debt in ML projects. The comparisons provided in Figure~\ref{fig.debt.comparison} are statistically significant with the Wilcoxon Signed Rank test $p$-value lower than 0.05 and a large Cohen's $d$ effect size difference.
When considering all the comments in ML and non-ML code from our dataset (without any sampling), 1.87\% of comment declarations in ML projects contain a technical debt admission (i.e., SATD), compared to only 0.93\% for non-ML projects, having a ratio of 2.02.
Our findings for SATD in non-ML software are aligned with the findings of prior research. For instance, Bavota and Russo~\cite{bavota2016large} found SATD in  0.3\% of the comments of non-ML Java-based projects, which is comparable to the 0.9\% obtained by our analysis for Python-based non-ML projects.

\begin{figure}[!hbt]
  \centering
  \includegraphics[width=0.6\columnwidth,keepaspectratio]{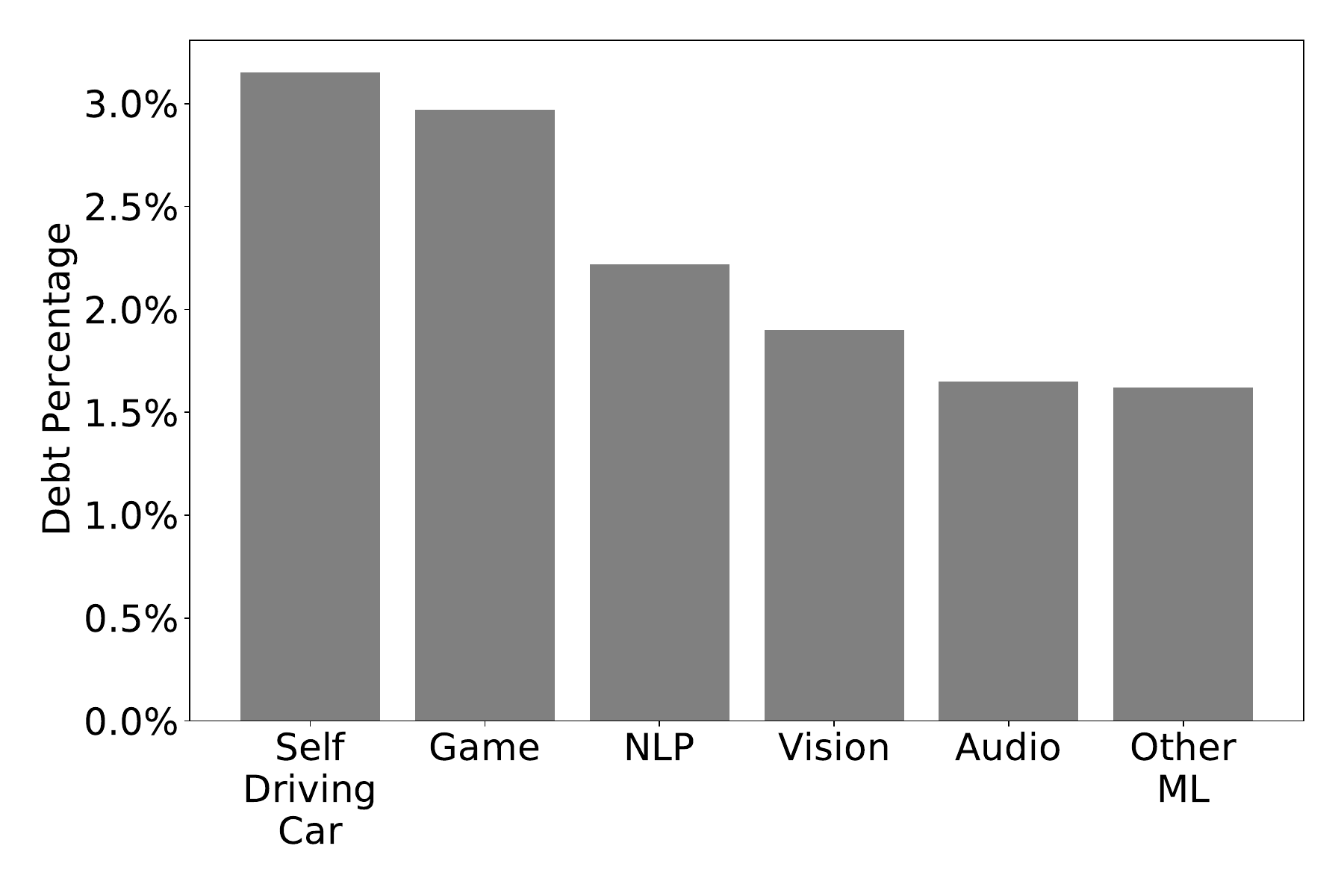}
  \caption{Distribution of SATD in the studied ML domains.}
  \vspace{-10pt}
  \label{fig.debt.domains}
\end{figure}

\textbf{We examined the prevalence of SATD across the five domains considered in our study and found that: 
} \textbf{Self driving car} and \textbf{Game} domains have higher proportions of SATD (i.e., 3.2\% and 3\% respectively) in comparison to the other domains. The \textbf{NLP, Image Processing, Audio processing} domains, and the \textbf{Other ML} category have  2.2\%, 1.9\%, 1.6\%, and 1.6\% of SATDs respectively. Figure~\ref{fig.debt.domains} presents the distribution of SATD across the five studied domains.

\begin{Summary}{Summary of RQ1}{}
Machine learning projects have a median percentage of SATD that is twice the median percentage of SATD in non-machine learning projects. 
\end{Summary}

\subsection{Discussion: Explanatory analysis for RQ1}\label{subsec.rq1.additionalDiscussion}

\begin{figure*}[!h]
    \centering
    \subfloat[Commit-level file changes normalized by LOC]{{\includegraphics[width=6cm]{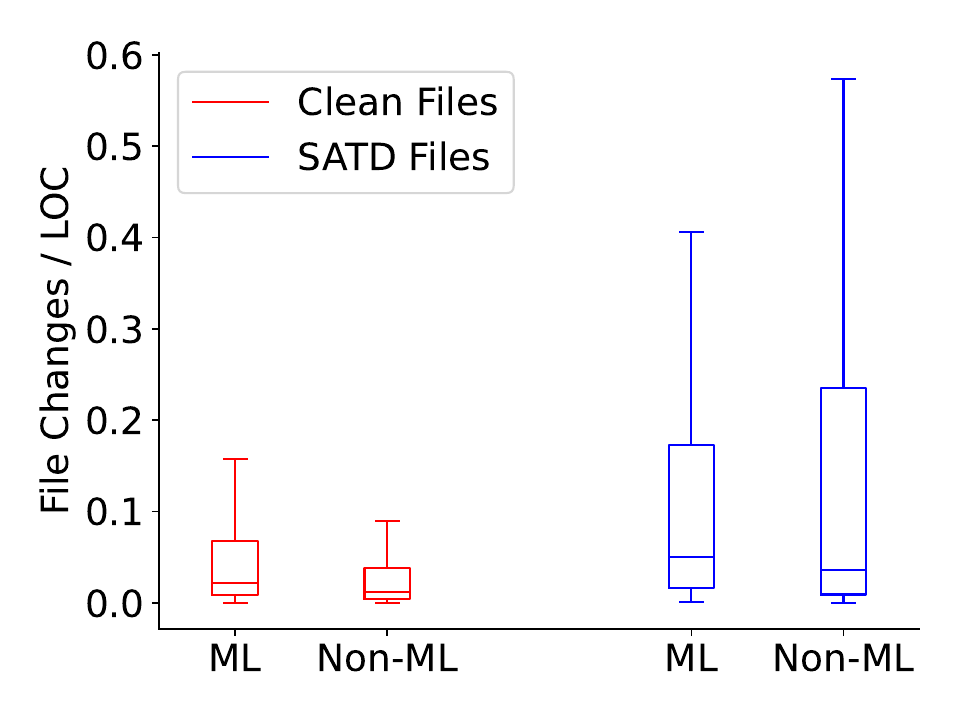} }}
    \qquad
    \subfloat[File-level complexity normalized by LOC]{{\includegraphics[width=6cm]{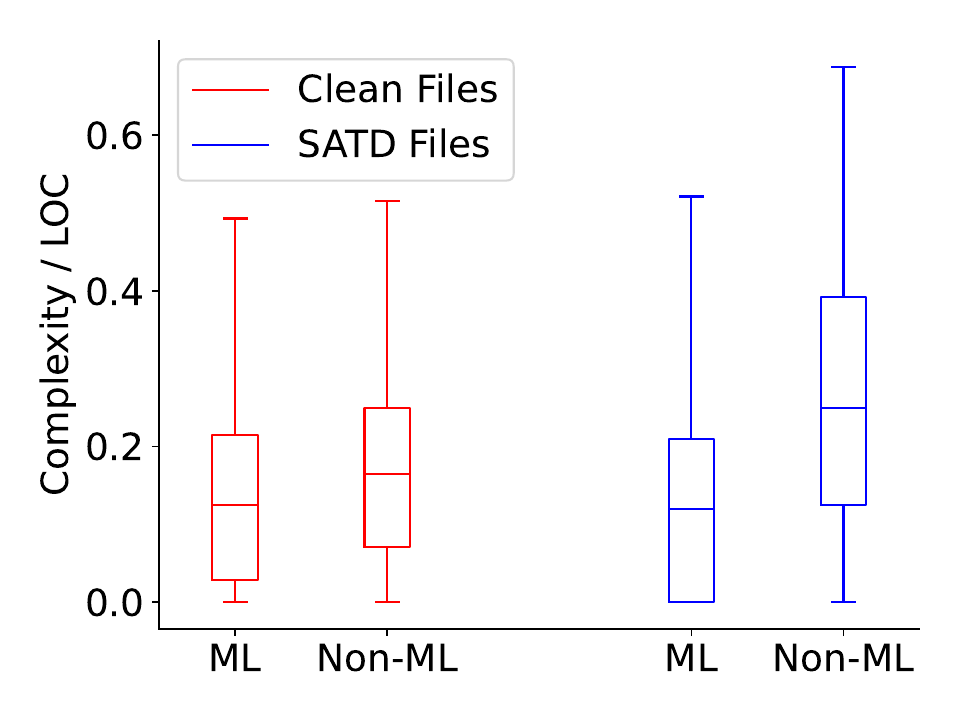} }}
    \caption{ML code has lower complexity, but more file changes than non-ML code.}
  \label{fig.complexity-f-changes}
\end{figure*}

\begin{figure*}[!ht]
    \centering
    \subfloat[Churn in ML and non-ML software]{{\includegraphics[width=6cm]{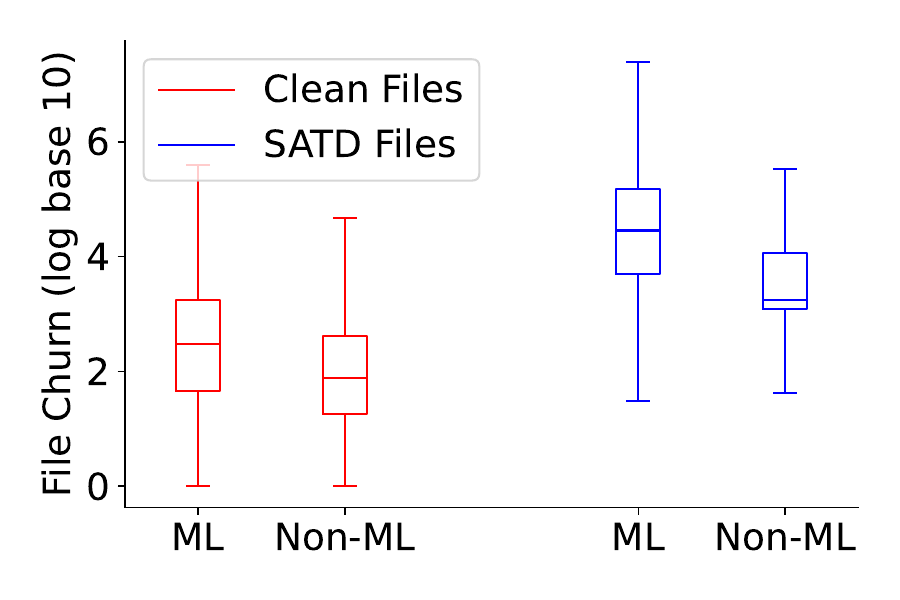} }}
    \qquad
    \subfloat[Churn-per-file size in LOC]{{\includegraphics[width=6cm]{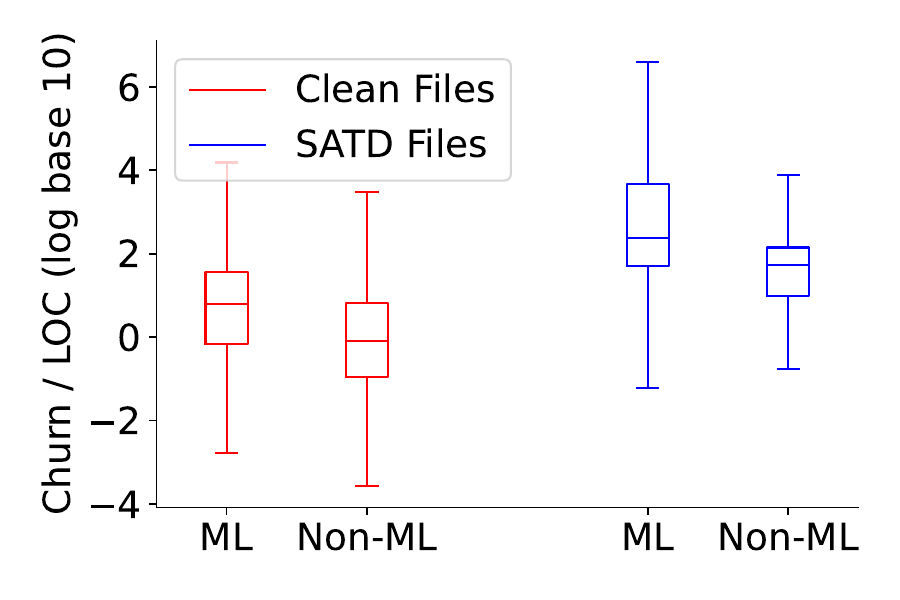} }}
    \caption{ML has much higher churn as compared to non-ML software projects.}
  \label{fig.churn}
\end{figure*}

Since ML developers tend to experiment frequently to find the optimal model, the ML codebase inherently undergoes significant modifications. Given that high code churn is widely recognized as a risk factor associated with poorer software quality and higher technical debt in general software engineering literature~\cite{mauerer2021search,Morales15}, we hypothesize it could similarly influence SATD prevalence in ML contexts. To investigate this explanatory hypothesis formally, we state the hypotheses:
\begin{tcolorbox}[colback=gray!10, colframe=gray!50, boxrule=0.3pt, arc=2pt, left=4pt, right=4pt, top=2pt, bottom=2pt]
\begin{small}
{H$_0$:} There is no significant correlation between code churn, file changes, file complexity, and the proportion of SATD in ML and non-ML projects.
\end{small}
\end{tcolorbox}

As mentioned in~\ref{subsec.approach.rq1}, we adopt Devil’s Advocate approach, individually testing each potential confound—churn, file changes, and complexity to see if any of them alone can explain away the ML vs non-ML SATD gap.

\textbf{Results.} Our analysis shows (Figure~\ref{fig.churn}(a)) that ML code indeed experiences significantly higher code churn compared to non-ML code (Wilcoxon Rank Sum test $p < 0.01$, Cliff's $\delta=0.8$), aligning with our expectation from the repetitive and iterative nature of ML development (e.g., continuous hyperparameter tuning and dataset updates). Additionally, SATD-containing files in ML code exhibit significantly higher churn normalized by file size (Figure~\ref{fig.churn}(b)), suggesting churn as a notable risk factor for SATD occurrence in ML contexts.

To test the devil's advocate, i.e., to rule out alternative explanations, we also analyze two other potential confounding factors—file-level complexity and the number of files changed per commit (diffusion). Contrary to churn, we find that file complexity is significantly lower in ML compared to non-ML projects (Figure~\ref{fig.complexity-f-changes}(b); Wilcoxon Rank Sum test $p < 0.01$, Cliff’s $\delta=0.2$ for SATD files), and the number of files changed per commit also shows negligible differences (Figure~\ref{fig.complexity-f-changes}(a); Cliff's $\delta < 0.1$). These results indicate that neither complexity nor the number of files changed per commit significantly explains the observed higher SATD prevalence in ML projects.

Overall, after explicitly controlling for potential confounding factors (code churn, file complexity, and file-change diffusion), the significantly higher prevalence of SATD in ML code persists. This strongly suggests that the high SATD density observed in ML projects is not merely an artifact of these structural metrics, but rather intrinsically linked to the specific nature and iterative development processes common in ML software development. ML developers should be aware of this intrinsic risk and proactively manage their technical debt, potentially through systematic monitoring of SATD indicators such as churn during model building or frequent hyperparameter updates.

\subsection{\rqtwo} \label{subsec.rq2}

\begin{table}[]
\caption{Distribution of different types of SATD in ML software. Types except for the bolded ones are reused from Bavota and Russo~\cite{bavota2016large}. The provided examples are hyperlinked to the source code file. 
}
\label{table.debt.taxonomy}

\begin{tabular}{|p{2.15cm}|p{3.1cm}|p{6.5cm}|p{0.8cm}|}
\hline
Type                         & Sub Type                    & Description [D] Example [E]                                                                                                                                                                                                           & \% age \\ \hline
\multirow{3}{*}{Requirement} & Functional \textgreater   
Improvement    & [D]: Comments reporting improvements to the existing implementation. [E]:\href{https://mypublicdoc.s3.us-east-2.amazonaws.com/DebtFiles/220\%23\%5EMachine Learning In Action\%5EChapter4-NaiveBayes\%5Evenv\%5ELib\%5Esite-packages\%5Esetuptools\%5Ecommand\%5Eeasy_install.py}{``Network Library could be added too''}                                            &       19.6    \\ \cline{2-4} 
                             & Functional \textgreater New Features   & [D]: Comments reporting missing features that need to be implemented later on. [E]: \href{https://mypublicdoc.s3.us-east-2.amazonaws.com/DebtFiles/319\%23\%5Eneuralmonkey\%5Eattention\%5Etransformer_cross_layer.py}{``TODO handle attention histories''}                                                                                                    &    15        \\ \cline{2-4} 
                             & Non Functional   \textgreater Performance           & [D]: SATD for improving the performance of the current implementation.  [E]: \href{https://mypublicdoc.s3.us-east-2.amazonaws.com/DebtFiles/131\%23\%5Esrc\%5Eencoder\%5Einference.py} {``TODO I think the slow loading of the encoder might have something to do with the device it''}                                                                       &      1.5      \\ \hline
\textbf{Configuration}       & -                           & [D]: SATD indicating implementation using unsure configurations. [E]: \href{https://mypublicdoc.s3.us-east-2.amazonaws.com/DebtFiles/173\%23\%5EHW3_Instance_Segmentation\%5Edetectron2\%5Etest.py} {``300 iterations seems good enough but you can certainly train longer'' }                                                                                   &      12.3      \\ \hline
\multirow{3}{*}{Code}        & Low Internal Quality        & [D] Reported SATD indicating issues with code quality.  [E]:\href{https://mypublicdoc.s3.us-east-2.amazonaws.com/DebtFiles/293\%23\%5Efarm\%5Einfer.py}{``TODO change format of formatted\_preds in QA list of dicts''}                                                                                                                                         &   10.9         \\ \cline{2-4} 
                             & Refactoring                 & [D] SATD indicating refactoring code or removing dead code. [E]:\href{https://mypublicdoc.s3.us-east-2.amazonaws.com/DebtFiles/319\%23\%5Eneuralmonkey\%5Edecoders\%5Etransformer.py}{``TODO make this code simpler''}                                                                                                                                          &       10.3     \\ \cline{2-4} 
                             & Workaround                  & [D] SATD incurring tentative or Ad-Hoc implementations which need to be fixed later. [E]:\href{https://mypublicdoc.s3.us-east-2.amazonaws.com/DebtFiles/70\%23\%5Emindmeld\%5Etokenizer.py}{``naive approach for generating forward mapping this is naive and probably not robust''}                                                                  &      9.4      \\ \hline
Design                       & Design Patterns             & [D]: SATD indicating violations of good Object-Oriented design. [E]:\href{https://mypublicdoc.s3.us-east-2.amazonaws.com/DebtFiles/202\%23\%5Eexperiment\%5Evenv\%5ELib\%5Esite-packages\%5Epip\%5E_vendor\%5Ehtml5lib\%5Etreebuilders\%5Ebase.py}{``XXX should this method be made more general''}                                                                   &    6.2   \\ \hline

Defect                       & Defects \textgreater Known Defect to fix         & [D]: SATD indicating known incorrect implementations leading to buggy behavior. [E]:\href{https://mypublicdoc.s3.us-east-2.amazonaws.com/DebtFiles/220\%23\%5EMachine Learning In Action\%5EChapter8-Regression\%5Evenv\%5ELib\%5Esite-packages\%5Epip-9.0.1-py3.6.egg\%5Epip\%5E_vendor\%5Edistlib\%5Eversion.py}{``TODO unintended side effect on e.g. 2003 05 09''} &       8.2     \\ \hline

Testing                      & \textbf{Inadequate Testing} & [D]: Reported SATD indicating missing tests for specific. [E]:\href{https://mypublicdoc.s3.us-east-2.amazonaws.com/DebtFiles/234\%23\%5Edeep-learning\%5ECapsNET\%5ETensorFlow_Implementation\%5EcapsLayer.py}{``TODO 1 Test the fully\_connected and conv2d function''}                                                                                      &    5.0   \\ \hline
Undefined                    &             -               & [D]: Comments indicated as SATD, but the debt description is missing. [E]:\href{https://mypublicdoc.s3.us-east-2.amazonaws.com/DebtFiles/74\%23\%5Edeeppavlov\%5Emodels\%5Ego_bot\%5Epolicy\%5Epolicy_network.py}{``todo"}
                             &      3.5      \\ \hline

\textbf{On-Hold}             &            -                & [D]: Existing SATD comments in the code even though the debt has been paid. [E]:\href{https://mypublicdoc.s3.us-east-2.amazonaws.com/DebtFiles/244\%23\%5Erun\%5Erelation_extraction\%5Emulti_head_selection\%5Edata_loader.py}{``todo maxlen'' (maximum length implementation was already done in code)}                                                        &      3.2  \\ \hline

Unknown                      &               -             & [D]: Debt unclear from the description. [E]:\href{https://mypublicdoc.s3.us-east-2.amazonaws.com/DebtFiles/206\%23\%5Edecision_tree\%5Edecision_tree(ID3_C4.5).py}{``TODO\begin{CJK*}{UTF8}{gbsn} 算法 \end{CJK*}4''}  
&      1.2    \\ \hline
Documentation                &              -           & [D]: SATD indicating missing documetation. [E]:\href{https://mypublicdoc.s3.us-east-2.amazonaws.com/DebtFiles/170\%23\%5Eweb.py}{``TODO add logger info''}                                                                                                                                                                                                    &      0.6    \\ \hline
\end{tabular}
\end{table}

\textbf{In addition to the traditional SATD taxonomy by Bavota and Russo, ML code faces two new types of SATDs, namely \textit{configuration debt} and \textit{inadequate tests debt}.} \textbf{The most frequent type of SATD in ML code is \emph{functional requirement debt}.} Table~\ref{table.debt.taxonomy} shows the distribution of SATDs observed in our studied sample. We found instances of \textit{code debt, design debt, requirement debt, and test debt}, which are identified in the taxonomy proposed by Bavota and Russo \cite{bavota2016large}. \emph{Functional requirements debt} accounts for 34\% of SATD in our sample. Among them, 19\% is related to features needing improvement (i.e., the SATD comment contains the statement ``improvement of features needed") and 15\% of \emph{functional requirement debt} is about new features (i.e., the SATD comment contain the statement ``new features to be implemented''). Code SATD accounts for a total of 30\% debt with \emph{low internal quality, refactoring} and \emph{workaround} SATDs, each accounting for 11\%, 10\% and 9\% of the instances respectively. \emph{Known defect to fix} accounts for 8\% of SATDs instances, while SATDs related to \emph{design patterns} account for 6.1\% of SATDs instances. We also identified eight instances (or 3\%) of \textit{On-Hold SATD}. On-Hold SATDs are comments in which developers explicitly document that implementation is paused pending an external condition or event (e.g., an open issue must be resolved or a dependency implemented elsewhere) \cite{maipradit2020wait, Rungroj}, yet the SATD comment remains in the code until that condition is met.
We identified two new categories of SATD not mentioned in the original taxonomy; i.e., \emph{configuration debt} and \emph{inadequate tests}, with 42 and 17 instances, respectively.

\begin{figure}[h]
  \centerline{\includegraphics[width=1.1\columnwidth,keepaspectratio]{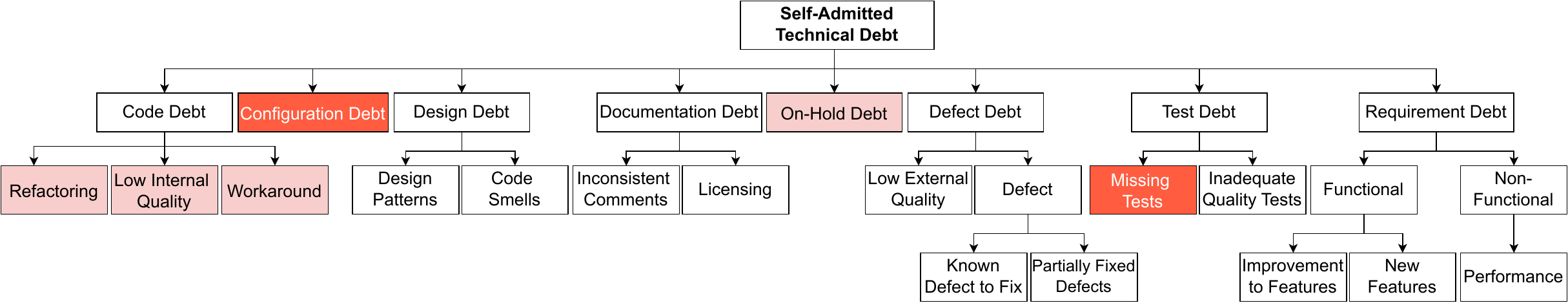}}
  \caption{Traditional and ML-specific SATD taxonomy.}
  \label{fig.debt_comprehensive_taxonomy}
\end{figure}

A visual representation of the debt taxonomy is provided in Figure~\ref{fig.debt_comprehensive_taxonomy}, where the uncolored debt types were not found in our research. The newly found ML debt types are highlighted in the darker color palette in Figure~\ref{fig.debt_comprehensive_taxonomy}, whereas the lighter orange colored ones are reused from Bavota et al.’s extended taxonomy~\cite{bavota2016large,Rungroj,suryanarayana2014refactoring}.

In our context, configuration debt refers to sub-optimal settings or lack of clear documentation on essential parameters governing ML system behavior, which includes, but is not limited to, algorithm hyperparameters, model settings, and data pipeline configurations. For instance, the comment ``\textit{TODO Also add batch\_size and other extraction specific configs to file}"\footnote{Accessed on: 2024-03-22, \url{https://github.com/Cloud-CV/visual-chatbot/tree/master//viscap/data/extract_features_maskrcnn.py}} signifies configuration debt as it alludes to an absence of standardized practices to track and manage essential configurations like batch sizes and extraction settings, potentially leading to inefficiencies and challenges in model reproducibility. Such configurations are needed to pre-process the data, as well as to tune the model (aka, hyper-parameters). Another example of such configuration debt is found in the project Image-Processing-via-deep-learning\footnote{Accessed on: 2024-03-22, \url{https://github.com/eddieczc/Image-Processing-via-deep-learning}}, where a developer mentioned that ``\textit{300 iterations seems good enough but you can certainly train longer}''. This SATD comment indicates the need to update hyperparameters' configuration. Similarly, a developer\footnote{Accessed on: 2024-03-22, \url{https://mypublicdoc.s3.us-east-2.amazonaws.com/DebtFiles/24\%23\%5Eviscap\%5Edata\%5Eextract_features_maskrcnn.py}} mentioned ``\textit{\# TODO: Also add batch\_size and other extraction specific configs to file}'', indicating improper management of configuration variables.\\
Regarding the \emph{inadequate test} SATDs, they occur when testing is postponed to the future. For instance, in project\footnote{Accessed on: 2024-03-22, \url{https://github.com/NVIDIA/DeepLearningExamples.git}}, developer(s) mentioned ``\textit{\# TODO(yonib): Also test logit\_scale with anchorwise=False}'' indicating missing / inadequate tests before committing the code. The taxonomy proposed by Bavota and Russo contains a ``Test'' category, but this category refers to the quality of test cases, while our new \emph{inadequate tests} category is about missing or inadequate testing against a functional aspect of the code. The absence of a necessary test case extends “test debt” in its traditional scope, that presumes the existence of some testing but of insufficient quality. This distinction highlights the severe impact of missing tests on the functionality and reliability of software, particularly in ML systems where such omissions can be particularly detrimental.

\begin{figure}[!ht]
  \centering
  \includegraphics[width=0.7\columnwidth,keepaspectratio]{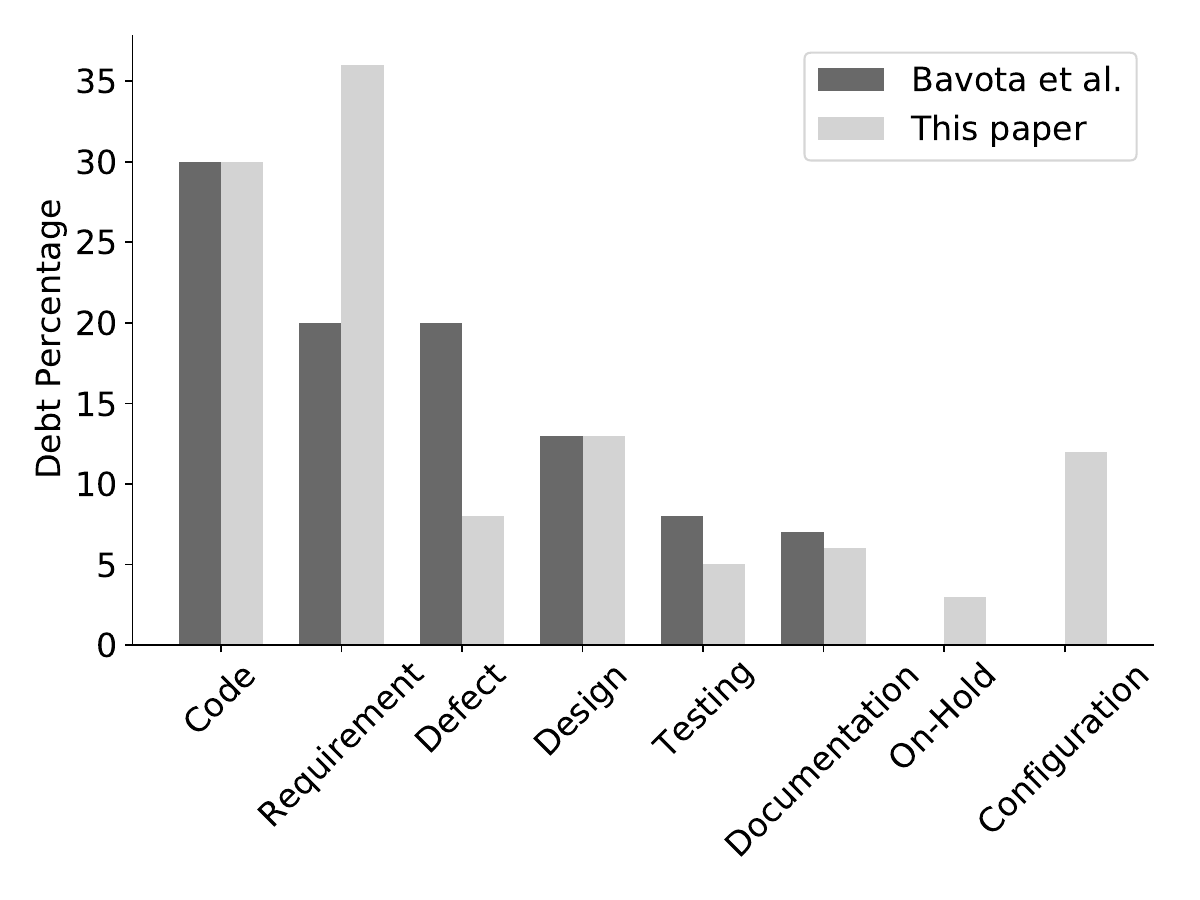}
  \caption{Comparison of types of debt identified by our research and prior research conducted by Bavota et al.~\cite{bavota2016large}. On-Hold and Configuration debts are unique to ML-based systems.}
  \label{fig.rq2.comparison}
\end{figure}

Bavota and Russo~\cite{bavota2016large} also found that instances of \emph{Code, Requirement, Defect, Design, Test, and Documentation debt} account for respectively, 30\%, 20\%, 20\%, 13\%, 8\%, and 7\% of SATD in traditional software projects. In this study, we found them to represent respectively 30\%, 36\%, 8\%, 13\%, 5\%, and 6\% of SATDs in ML software projects (as shown in Figure~\ref{fig.rq2.comparison}). This indicates that ML software has lower proportions of defect and design debts and higher proportions of requirement debt than traditional (non-ML) software projects. However, the proportions of code, test, and documentation debts are similar, while we also found new SATD types of Configuration debt (12\%) and Inadequate tests (8\%).

\begin{Summary}{Summary of RQ2}{}
The categories of SATDs that dominate in ML projects are Requirement debt (36\%), Code (30\%), Configuration debt (12\%), Known defect to fix (8\%), and Inadequate tests (5\%) debts.
 \textit{Configuration debt} and \textit{Inadequate tests} were not found in traditional software projects. 
\end{Summary}

\subsection{\rqthree} \label{subsec.rq3}
\begin{table*}[!h]
\caption{Results of manual analysis of SATD found in the different stages of machine learning code. The Table rows have been ordered in decreasing order of percentage of occurrence. The examples are hyperlinked to the source code.}

\begin{tabular}{|p{2.5cm}|p{5.5cm}|p{3.5cm}|p{1cm}|}
\hline
\textbf{ML stage} &   \textbf{Description} & \textbf{Example} & \textbf{\% age}\\ \hline

Model Building &  Training the model including computation,
optimization of the cost function, and hyperparameter tuning. 
& \href{https://github.com/erikbern/deep-pink/blob/master/train.py}{``Train on a negative log likelihood of  classifying the right move''}
&  29.0\% \\ \hline

Data Preprocessing &
Transforming data into features or tensors that can be consumed by the model for its training and validation purposes.
& \href{https://github.com/bfelbo/DeepMoji/blob/master/deepmoji/filter_utils.py}{``ugly hack handling non-breaking space no matter how badly its been encoded in the input''}
&  24.7\% \\ \hline

Data Reading & 
Reading data from external sources like databases, csv, stream/batch processes.
&  \href{https://github.com/pannous/tensorflow-speech-recognition/blob/master/speech_data.py}{``TODO! see github url for some data sources''} 
& 8.6\% \\ \hline

Model Deployment & 
Deploying the ML pipeline for application-specific use.
&  \href{https://github.com/hzy46/Deep-Learning-21-Examples/blob/master/chapter_5/research/slim/datasets/process_bounding_boxes.py}{``Note: There is a slight bug in the bounding box annotation data.''}
& 7.5\%  \\ \hline

Model Validation & 
Validation of the correctness of the model and testing its robustness. 
&  \href{https://github.com/google-research/morph-net/blob/master/morph_net/framework/op_regularizer_manager_test.py}{`Check that regularizers were grouped properly.''}
& 4.0\% \\ \hline

\end{tabular}
\label{tab.Debt.dist}
\end{table*}

\textbf{The majority of SATDs in ML code resides in \emph{Model building} and \emph{Data Preprocessing} stages}. 
We found 111 SATDs belonging to \emph{Model Generation} and 92 SATDs in \emph{Data Preprocessing}. We also found 69 SATDs within the non-ML code portion of the studied projects. Finally, we found 15 SATDs in \emph{Model Validation}, 29 SATDs in \emph{Model Deployment} and 32 SATDs in the \emph{Data Reading}. We could not identify a corresponding stage for 22 SATD samples. 
We present examples of SATD found in the different ML stages in Table~\ref{tab.Debt.dist}, alongside the description of the stage and the proportion of SATDs found in the stage.
For instance, the following SATD comment ``\textit{TODO: Keras Batch Normalization mistakenly refers to var}" in the project ``YAD2K: Yet Another Darknet 2Keras''\footnote{Accessed on: 2024-03-22, \url{https://github.com/farzaa/DeepLeague/blob/master/YAD2K/yad2k.py}Line:144} belongs to the Model Training stage.

\begin{figure}[!h]
  \centering
  \includegraphics[width=0.7\columnwidth,keepaspectratio]{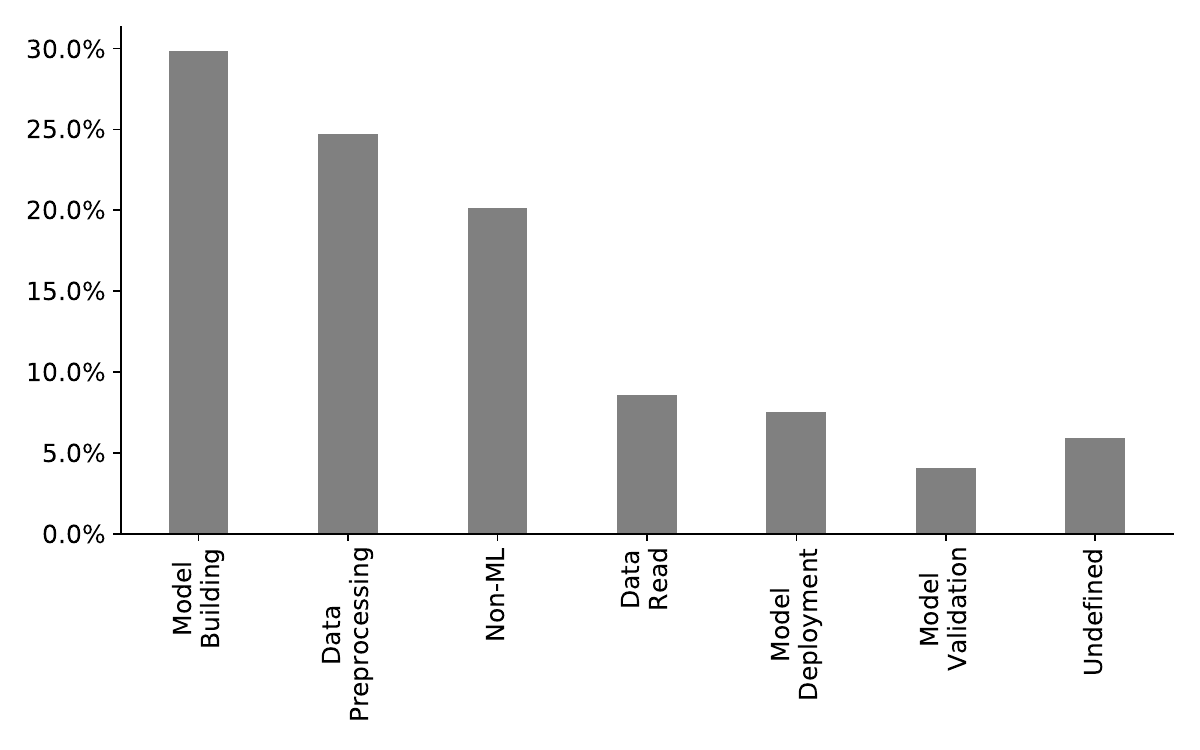}
  \caption{Percentage of the analyzed SATD across the different ML stages.}
  \label{fig.debt.taxonomyRQ3}
\end{figure}

\textbf{Model Building has the largest proportion of SATDs (46.7\%), 
whereas Model Validation has the lowest, i.e., 0.4\%.} This result is not very surprising given the amount of code churn that the Model Building stage experiences, as a result of the frequent retraining that is required to cope with concept drift issues. 

\begin{Summary}{Summary of RQ3}{}
Overall, SATD occurs in all five key stages of ML software, i.e., Data Reading, Model Training, Feature Transformation, Model Validation, and Model Deployment. However, Model Training has the largest proportion of SATD, whereas Model Validation has the lowest proportion. 
\end{Summary}

\subsection{\rqfour} \label{subsec.rq4}
\begin{figure}[!h]
  \centering
    \includegraphics[width=0.7\columnwidth,keepaspectratio]{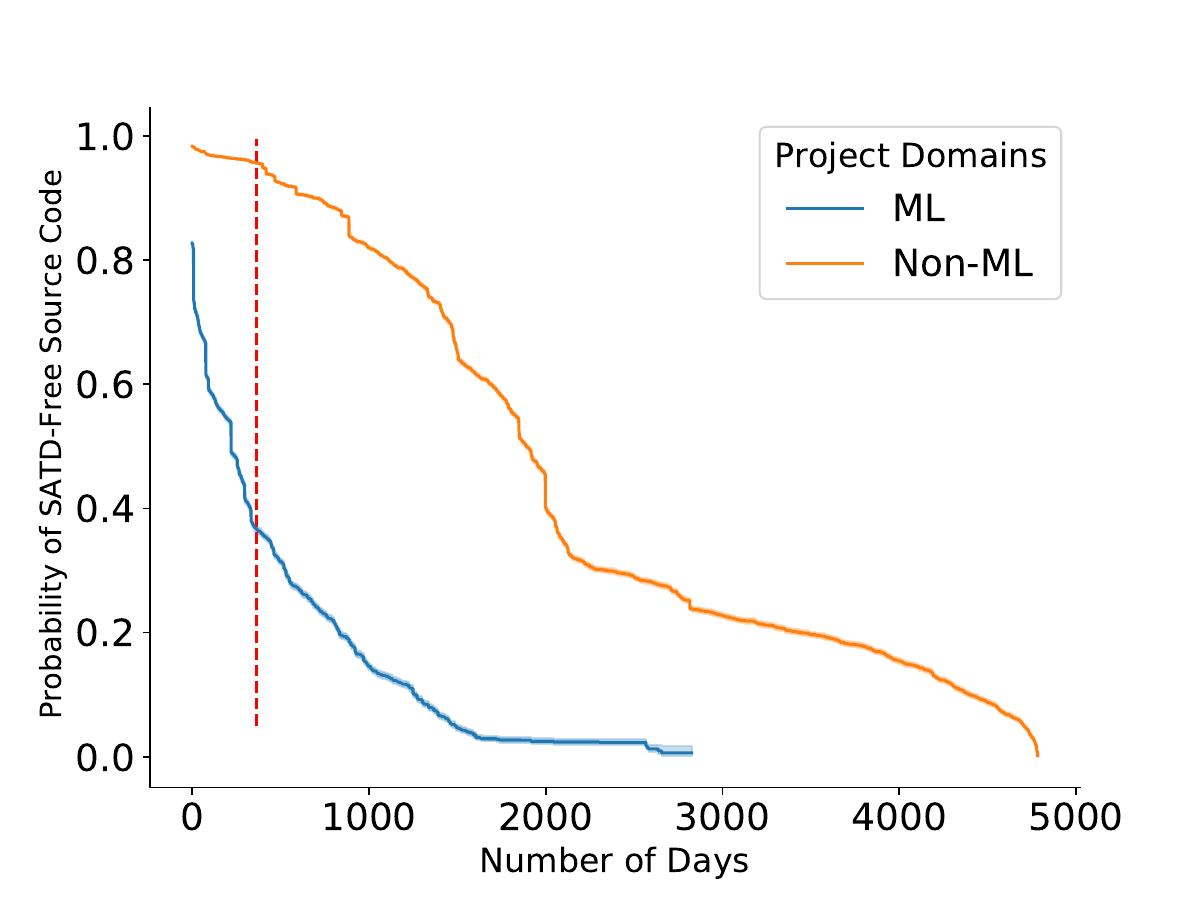}
  \caption{Survival analysis for the introduction of SATD into a clean source code file for ML and non-ML projects. The dotted red line represents the 1-year mark with 0.95 and 0.37 probabilities of survival of a clean source code file in non-ML and ML projects respectively.}
  \label{fig.survival.add}
  \vspace{-10pt}
\end{figure}

\textbf{\\Debt addition.}
\noindent{\bf SATDs are introduced in 140 times earlier during the development process of the studied ML projects in comparison with non-ML projects.} A median ML debt is added in 10 days, while a median non-ML debt is added in 1,405 days. Figure~\ref{fig.survival.add} shows the survival curves in ML and non-ML code, with the difference between both curves is statistically significant with Wilcoxon rank sum test p-value=0.0. The steep decline of the ML curve in Figure~\ref{fig.survival.add} reflects a faster decline of the probability of ML projects to remain clean, i.e., remain without SATD. Within just one year of development time, ML projects have a low probability (i.e., 0.36) of its source code remaining debt-free, whereas non-ML projects have a much higher probability (i.e., 0.9). Even after 4 years, the probability of SATD introduction in non-ML projects is still below 0.3 (i.e., probability of SATD-free source code above 0.7), while it is 0.05 for ML projects.

\begin{figure}[!h]
  \centering
  \includegraphics[width=0.7\columnwidth,keepaspectratio]{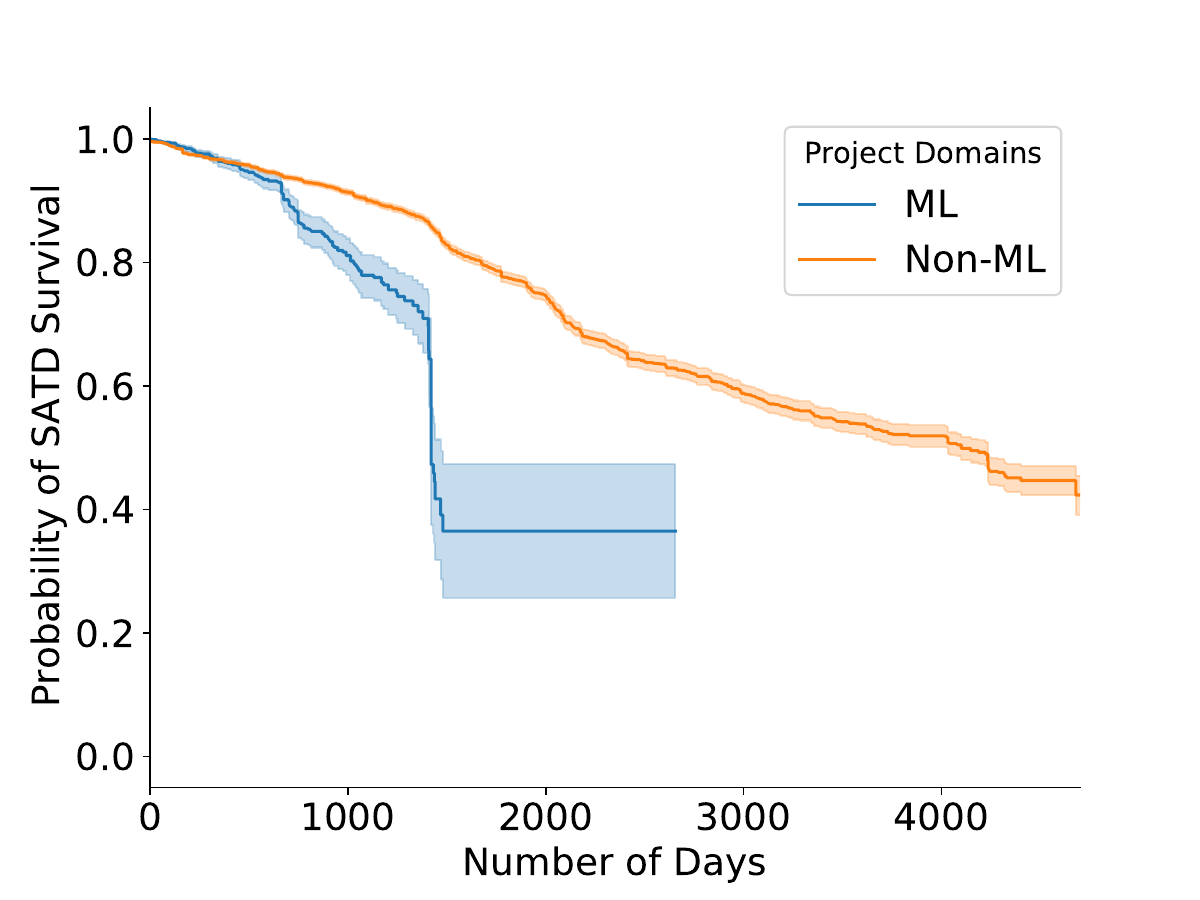}
  \caption{Survival analysis of existing SATD in ML and non-ML source code files.}
  \label{fig.survival.rem}
  \vspace{-10pt}
\end{figure}

\textbf{\\Debt removal.}
\textbf{The survival analysis indicates 3.7 times faster SATD removal in ML projects in comparison with non-ML projects in our studied dataset.} After its addition, a median ML SATD is removed in 1.1 years, whereas a non-ML SATD is removed in 4.1 years. Figure~\ref{fig.survival.rem} shows the likelihood of debt removal in ML and non-ML projects. After a two-year time span, the probability of SATD to survive in ML projects is 0.89, while this probability is 0.94 for non-ML projects. Such observations imply that ML developers may be cognizant of the faster debt accumulation in ML code and hence dedicate more time and effort towards its mitigation than developers working on non-ML code.\\

While our findings corroborate the idea that SATD in ML projects is both introduced and resolved more rapidly than in non-ML projects, it is important to put these numbers in context. Alternative reasons include 1) shorter lifetime of SATD in ML projects may be due to the shorter life cycle of these projects. In fact, in our data set, the medium lifecycle of ML projects is 1.15 years, while it is 4.16 years for non-ML projects. This is expected, as ML software was relatively new when the data for this study was collected; 2) the repetitive nature of ML experiments may flush out debt while dead experimental code paths are being refactored. This is corroborated by the 52\% of cases of SATD removal in ML code that coincided with the deletion of the actual source code file. For the remaining 48\% of cases, the file was only modified. This may indicate that SATD removal often happens during massive refactoring, where the debt-containing files are deleted. 3) the high code churn characteristic of ML projects, reflecting their fast-paced and iterative development cycles. This higher code churn rate in ML software development was further discussed and evidenced in the results of RQ1 (see Section~\ref{subsec.rq1}) and Figure (\ref{fig.ml_non_ml_comparison}-d), providing additional support for our initial assertions. Furthermore, prior research found that higher churn also correlates with higher likelihood of bugs~\cite{Morales15,mauerer2021search}. \\
Overall, further research, particularly qualitative studies are needed to determine whether rapid SATD turnover in ML projects reflects healthy refactoring or risk-prone instability.


\begin{Summary}{Summary of RQ4}{}
SATD is both introduced and removed significantly earlier in ML projects in comparison to non-ML projects. 
\end{Summary}

\subsection{\rqfive} \label{subsec.rq5}

\begin{figure}[!h]
  \centering
  \includegraphics[width=0.7\columnwidth,keepaspectratio]{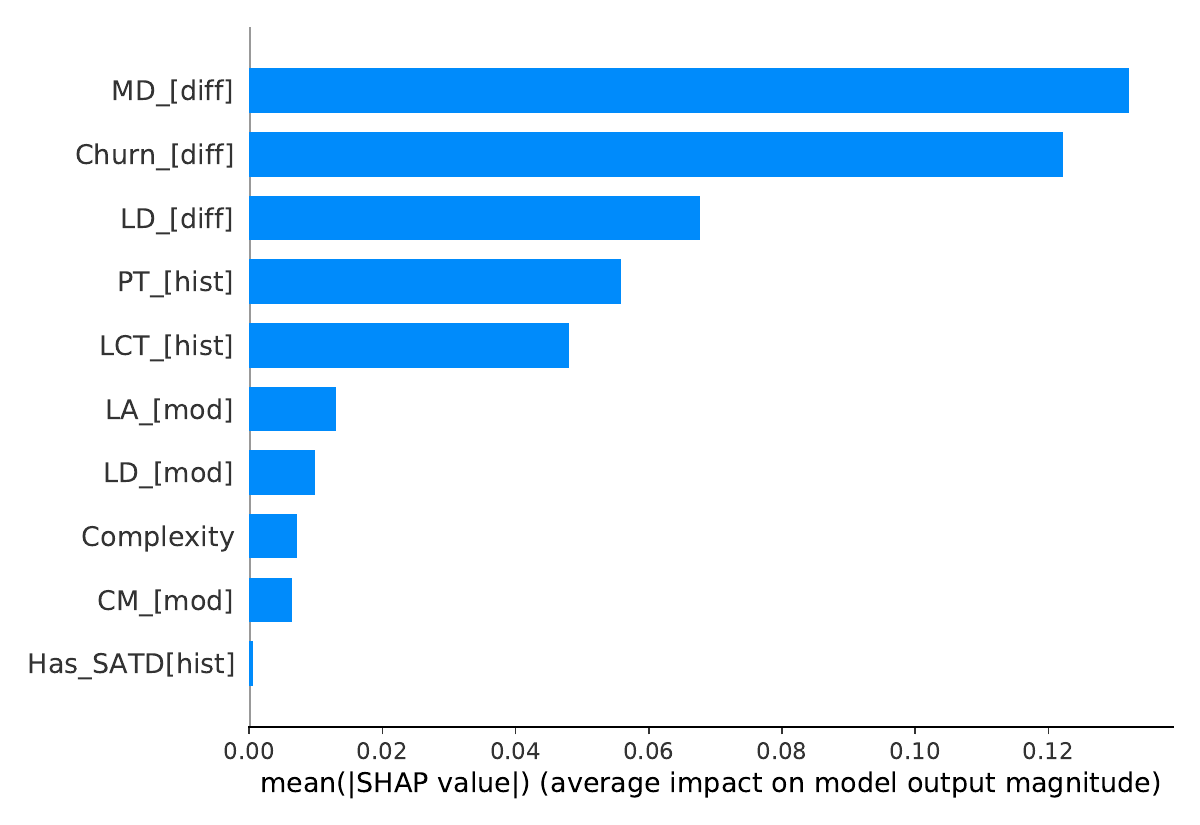}\caption{
  Global model interpretation from SHAP for prediction of long lasting SATD.
}\label{fig.interpretation}
\end{figure}

Figure~\ref{fig.interpretation} shows mean Shapely values obtained for the best performing Random Forest model (from the bootstrapping analysis). Our models from the bootstrapping analysis have a median Precision of 93\%, Recall of 98.8\%, and F1 Score of 95.8\%. \\
\textbf{SATD introduced during large code changes (Churn\_{[}diff{]}) spanning multiple files (MD\_{[}diff{]}) are likely to be long-lasting SATD} according to our explanatory model. The timing of the commit, i.e., PT\_{[}hist{]} (Project Time) and LCT\_{[}hist{]} (Last Commit Time) also appear to be important factors in the duration of SATDs in the studied projects. Long-lasting SATDs are added at a median of 231 days, as compared to quickly removed SATDs which are added at a median of 7 days (i.e., towards the beginning of the project). 

Interestingly, long-lasting SATDs are added in smaller files. The median complexity of files containing long-lasting SATDs is 25, while the median complexity of files where SATDs are removed quicker is 40. This finding suggests that ML developers pay special attention to SATDs occurring in complex files and tend to ignore those scattered through low complexity files. 

One hypothesis is that when SATD is introduced into small files during commits that introduce large changes, developers may perhaps orient their attention towards the files that perform more functionality (high complexity), thereby overlooking correcting the small ones. Such debt may be deemed non-urgent by the developers. Since the functionality of such low complexity files is trivial, debt in these files may be considered ``harmless”, and hence debt correction is not prioritized as it is for the debt instances in large files. 

Adding to the above hypothesis, it is also probable that resource allocation might be a factor, with developers focusing their efforts on managing SATD in larger, more critical files, similar to how testing larger files might find more bugs~\cite{mcintosh2014impact,zhang2012empirical}. Risk perception could also play a part, as SATD in smaller files might be viewed as less likely to cause major issues. Familiarity might contribute too, as developers are potentially more at ease managing SATD in larger files they frequently work on. Oversight during code reviews, due to the less complex nature of small files, might allow more SATD to persist. Developers could also underestimate the impact of SATD in these files, falsely assuming its repercussions will be negligible. Additionally, it is plausible that smaller files, deemed easier to rewrite or refactor in the future, might have their SATD issues postponed. Lastly, the cognitive load to understand and fix SATD in larger files might draw developers' focus away from smaller ones. Future work should validate these hypotheses.

\begin{Summary}{Summary of RQ5}{}
Long-lasting SATDs are introduced during large code changes spanning multiple files (with low complexity). Long-lasting SATDs are added at a median of 231 days as compared to quickly removed SATDs which are added at a median of 7 days (i.e., towards the beginning of the project). 
\end{Summary}

\section{Threats to Validity}\label{sec.threats}
\subsection{Internal Validity}
The act of ``self-admitting'' technical debt can manifest through various channels, including GitHub issues or project management tools such as Trello and Jira. Consequently, measuring SATD solely through code comments might not encompass the entire spectrum of "self-admission" of debt in software projects. Nevertheless, Potdar and Shihab~\cite{potdar2014exploratory} have highlighted SATD in code comments as a discernible metric for (SA)TD. Furthermore, we follow authoritative research (\cite{liu2018satd, Zampetti, liu2020using, alves2014towards, maldonado2015detecting, bavota2016large, da2017using, wattanakriengkrai2018identifying, liu2020using, kruchten2012technical, kruchten2013technical}) in the area of self-admitted TD, that have estimated SATD via code comments.

The SATD identification tool used in our study may provide false positives. Although there are many SATD identification tools, we chose a NLP-based SATD detector proposed by Liu et al.~\cite{liu2018satd}. This tool, unlike other state-of-the-art debt detector tools that provide high precision and low recall, has a high recall. Moreover, this tool has been utilized in several previous studies, such as those by Liu et al.~\cite{liu2021exploratory} and Zampetti et al.~\cite{Zampetti}.

In addressing \textbf{RQ2} and \textbf{RQ3}, two of the authors manually examined 611 instances of SATDs. This process may have introduced a subjective bias. However, given the high Cohen Kappa score of 94\% between the two authors, we believe that the risk of bias is minimal. Furthermore, to compare the qualitative results of \textbf{RQ2}, we rely on the findings of Bavota and Russo~\cite{bavota2016large} and do not use our sample of 318 non-ML projects. This is because our extensive qualitative analysis, which took several months of efforts to label the 611 SATD samples, is non-trivial. Replicating such efforts on a sample of non-ML SATD instances was not possible in the scope of this study. 

\subsection{Construct Validity}
There is no consensus among researchers regarding the inclusion of missed requirements or known defects in the concept of SATD. The seminal work by Kruchten et al.\cite{kruchten2013technical} does not regard requirement and defect debt as SATD, however, the work of Bavota et al.~\cite{bavota2016large} does consider the two categories as SATD. Despite these differences, as the majority of research (\cite{alves2014towards, maldonado2015detecting, bavota2016large, da2017using, wattanakriengkrai2018identifying, liu2020using}) on SATD accepts Requirement and Defect Debt as SATD, we include these categories in our manual analysis for \textbf{RQ2}. This potential variation in the understanding and interpretation of SATD poses a construct validity threat as it could lead to varying interpretations and thus results.

The curation of ML repositories involved two levels, with an initial level using programmatic checks to verify criteria \textit{C1, C2}, and \textit{C3}, and with level-2 using manual checks against criteria \textit{C4} and \textit{C5}. Our Level-2 checks were done by one author, which represents a minor threat to validity due to subjective bias. We believe that the threat is minimal since for both ML and non-ML repositories, the first step is to check the project’s Readme itself for the criteria \textit{C4} and \textit{C5}, which is a straightforward process. Typically, README files in software development include sections on the project's purpose, installation instructions, usage examples, etc.~\cite{prana2019categorizing}. While a programmatic string-matching-based tool could potentially scan for ML-related keywords (eliminating the need for a manual analysis altogether), our manual efforts were only done to make the process more resilient to errors. 
Furthermore, checking for criteria \textit{C4} and \textit{C5} using a highly structured approach as outlined in sub-section 4.1.3 is a curation process, inherently objective in nature, which can minimize subjective bias~\cite{hallowell2009techniques}. This manual curation for repository selection is opposed to the qualitative analysis done for \textbf{RQ2} and \textbf{RQ3}, where more substantial qualitative coding is required, for which we performed multi-rater coding with high inter-rater agreements.

In this work, we use survival analysis to examine for each project the time until the introduction of SATD. Specifically, we use the non-parametric Kaplan-Meier estimator. We also used the non-parametric Wilcoxon Rank Sum test to compare populations. Both these tests do not require the data to be normally distributed. In \textbf{RQ5}, we assume that the admitted technical debt is fixed when the SATD comment is removed. However, it could be possible that the debt still lingers on even after removing the SATD comment.

About 40\% of SATD instances were categorized as `Undefined', `Unknown', or `False Positives', reflecting the subjective nature of this classification in RQ2. This ambiguity, particularly in distinguishing between actual technical debt and inaccurately flagged instances may potentially affect our RQ1 findings. 
However, given the inherent uncertainties and challenges associated with categorizing SATD, and it is reasonable that both ML and non-ML SATD exhibit similar patterns of being undefined, unknown or the tool detecting false positives. Hence, our main conclusion regarding higher prevalence of SATD in ML projects still stands, and the “potential” parallel does not undermine our core conclusion of RQ1, i.e., ML debt is twice more prevalent than non-ML debt.

Finally, for \textbf{RQ4} and \textbf{RQ5}, we limited our extraction of SATD to code modifications solely from the master branch. It is worth noting that SATDs might have been introduced or removed in branches other than the master branch. Our decision to focus on the master branch was due to the current challenges in constructing a change log post-rebasing — an unresolved issue that remains a topic of active research. Including data from other branches would not only expand our dataset substantially but could also introduce complexities in accurately mapping the introduction and removal of SATDs across different branches. Moreover, the master branch is the ``final'' code meant to be seen by the rest of the organization. Hence, SATD in the master branch is of high importance and impacts others who are using the repository.

\subsection{External Validity}
The results of our findings are dependent on our sampled data. To compare ML and Non-ML Projects in \textbf{RQ1} and \textbf{RQ4}, we have taken steps to ensure that the selection process is consistent for both ML and Non-ML fronts. While the sample of ML projects was based on exhaustive search (i.e., including relevant projects in a domain until subsequent projects are irrelevant), procuring non-ML projects was based on convenience sampling (getting the top-318 non-ML projects as long as the selection criteria are met). Although a random sample would offer a more unbiased comparison, obtaining all non-ML Python projects across all GitHub pages, then selecting a random sample may not be feasible.

Moreover, although we select mature projects by using selection criteria inspired by related work~\cite{kalliamvakou2014promises,chen2020studying,alfadel2021use} such as a project requiring more than a month of development history, it is still possible that immature projects may seep into our analysis, biasing our results. Following prior research~\cite{alfadel2021use,chen2020studying} that has commonly mitigated this threat using threshold-based criteria, we establish five filtering criteria (e.g., a project should have more than one month of developmental history) for the removal of toy projects. 
Notably, our ML dataset lacks test cases. This absence signifies a deviation from traditional Software Engineering where tests involve unit tests and integration tests; and code quality is measured through the coverage of such tests. However, such testing practices do not seamlessly translate to ML scenarios~\cite{zhang2020machine}\footnote{Accessed on: 2024-03-22, \url{https://fg-tav.gi.de/fileadmin/FG/TAV/47.TAV/TAV47-Felderer-TestingTheUntestable.pdf}}\footnote{Accessed on: 2024-03-22, \url{https://shorturl.at/dnBDU}}. 
Instead of test cases, ML testing often involves ensuring that a model has high performance, high robustness, lack of bias, etc. Indeed, the lack of quality assurance standards (for instance \% test coverage which is easily measured in traditional software) in ML systems, in itself is a motivation for our study.

We conducted an analysis of a diverse portfolio consisting of 318 projects from well-established ML domains like ``self driving car'' and ``natural language processing'' to ensure that our findings are representative of industry-grade ML projects. However, our results may not encapsulate the SATD phenomena across the entire spectrum of ML software systems.

In Software Engineering research, it is widely acknowledged that datasets inherently exhibit diversity in terms of project size, versioning, and other attributes. This diversity is not viewed as a flaw but rather as a reflection of the inherent variability of software projects. As such, the fact that our filtering of the data led to slightly varying numbers of data points across the different ML repositories is not uncommon in empirical studies. Instead of (1) downsampling all ML domains to the size of the smallest domain, which would lead to a large loss of valid data points, or of (2) aggregating smaller domains to obtain fewer, but larger domains, which would introduce imprecision and noise, we opted to keep the ML domains as is, with heterogeneous sizes.

In \textbf{RQ3}, our analysis was confined to the examination of SATD within five stages (i.e., \textit{Model Building, Data Prepossessing, Data Reading, Model Validation,} and \textit{Model Deployment}) of the ML pipeline as proposed by Amershi et al. \cite{amershi2019software}. Consequently, the identified technical debt may not fully represent the nuances of TD that could occur in expanded areas of ML pipelines like model monitoring, maintenance, and evolution. Future studies could aim to analyze TD in these expanded areas or in MLOps development and monitoring tasks to provide a more comprehensive understanding of TD in overall ML-based software development. This limitation suggests that our results are primarily applicable to the specific context we investigated, and caution should be taken when generalizing the findings to these other contexts.

\subsection{Conclusion Validity}
Just like debt in monetary situations, it may be argued that not all technical debt is bad and needs to be paid off eventually. Technical debt (TD) manifests in various forms, such as organizational, discussion, and product-related TD, and not all TD is detectable by tools. Prior studies by Potdar and Shihab~\cite{potdar2014exploratory} recognized SATD as an effective heuristic for estimating TD in software systems. The importance of understanding SATD for software maintenance and management has been acknowledged in previous non-ML research~\cite{Li2015, Alves2016, Rios2018, Alfayez2020, Lim2012, wehaibi2016examining, brown2010managing, letouzey2012managing, kruchten2012technical, kruchten2013technical, bavota2016large}, and generally the prevalence of SATD is considered a bad code smell~\cite{fowler2018refactoring}.

While our research does not conclusively measure the effort involved in debt removal and in estimating its crippling effects, we lay a strong foundation for future investigations. Notably, we found that most of the debt exists in ML model building stages (as found in RQ3), and the most recurring debt is ``improvement of functional requirements" (as found in RQ2). This finding is supported by our identification of a new debt type, “configuration debt”, indicating sub-optimal implementations in data preprocessing and model building. Developers speculating on sub-optimal ML pipelines can have an even higher cost than traditional software due to the tight coupling between different stages of the ML pipeline. TD in ML can have far-reaching detrimental effects across the pipeline, causing inefficiencies in the training process or ML model selection process to increase maintenance efforts. 

Overall, understanding the dynamics of SATD in ML software is essential for both software researchers and ML stakeholders. As our research shows, managing TD effectively in this area can significantly impact the overall quality and efficiency of ML software.

\section{Implications and Recommendations for ML Stakeholders}\label{sec.implications}
Our RQ1 evidence shows that ML repositories accumulate SATD twice as fast as comparable non-ML code. Debt clusters in two main hotspots: model-building (47\%) and data-pre-processing ($\approx$ 25\%), and are dominated by functional-requirement debt (34\%), configuration debt (12\%), and the newly observed inadequate testing debt (5\%). Based on these specific patterns, we recommend implications for practitioners, researchers, toolsmiths, and educators.

\textbf{1. Address configuration debt early}
For ML software, configuration debt is one of the prominent ML-type debt. Some prominent examples from our analysis include ``\# hardcoded search strategy''\footnote{last accessed Apr 17, 2025 \url{https://github.com/chris-chris/pysc2-examples/blob/master/a2c/kfac.py}}, or ``\# TODO test with different speeds''\footnote{last accessed Apr 17, 2025 \url{https://github.com/NVIDIA/DeepLearningExamples/blob/master/CUDA-Optimized/FastSpeech/generate.py}} or ``\# TBD: use better initializers (uniform, etc.)''\footnote{last accessed Apr 17, 2025 \url{https://github.com/kermitt2/delft/blob/master/delft/sequenceLabelling/preprocess.py}} show notable examples of configuration debt that is likely to accumulate in ML systems. 
An actionable item for ML practitioners would be to remain vigilant towards the accumulation of configuration debt by ensuring that all hyperparameter settings and experimental configurations are systematically tracked. This can be achieved through the use of experiment-tracking tools such as DVC or MLflow, where configurations are stored in version-controlled YAML or JSON files rather than being hardcoded within the source code~\cite{sklavenitis2025scoping}. 
To enforce this, pre-commit hooks or continuous integration (CI) pipelines can be employed to automatically reject code submissions that introduce ad-hoc constants or fail to reference external configuration files. Moreover, each configuration file can be tagged with an assigned owner responsible for its maintenance, along with a predefined review date or expiry period. Upon reaching this date, CI systems can be configured to automatically generate review tasks or issues, prompting the team to reassess the relevance and accuracy of the configuration. Such practices not only minimize the accumulation of configuration debt but also enhance reproducibility and maintainability within ML systems~\cite{sklavenitis2025scoping}. Toolsmiths can build CI tools that have ability to enforce such settings. \\

\textbf{2. Align Technical Debt Remediation times (budget payment windows) with High-Churn Development Phases}
Our survival analysis (RQ4) shows that SATD in ML code typically appears within 10 days of file creation, compared to 1,400 days in non-ML code. Additionally, our findings (RQ5) indicate that SATD often arises during commits that modify many files but exhibit low per-file complexity, suggesting rapid, large-scale changes with limited design consideration. These patterns point to quick, exploratory development cycles as a key contributor to persistent technical debt.

To mitigate this, we recommend that \textit{ML practitioners} should proactively schedule targeted refactoring shortly after such high-churn activities. Version control systems or CI pipelines can specifically monitor the ratio of normalized churn (i.e., churn to lines of code), teams can identify commits in the top quartile of normalized churn, and flag such commits for careful review/refactoring. Later in the budget friendly periods of downtime where delivery pressure is manageable, debt can be repaid with a special focus on high churn files, and those that were changed in alarming commits could be reviewed for potential SATD and considered for refactoring in the immediate following sprint. This approach aligns with our empirical observation that timely intervention can reduce the likelihood of SATD persisting long-term by approximately 20\%. These strategies can systematically keep technical debt in check for the long-run, improving maintainability without requiring disruptive, large-scale overhauls.\\

\textbf{3. Prioritize Rigorous Testing for Model-Building Components.}
Our study identifies a category of technical debt more-likely found in ML systems: \texttt{inadequate testing debt}. These are developer comments acknowledging missing or insufficient tests in critical parts of the ML pipeline, especially in model-building code (e.g., “TODO 1 Test the fully\_connected and conv2d function”). Such gaps can lead to silent failures or degraded model performance, as untested ML components may behave unpredictably under certain inputs or could lead to ill-defined configurations, causing unwanted behaviours (like low accuracy) in models or leading to unstable model training~\cite{de2025classification,sklavenitis2024measuring}. 

To address this, we recommend that ML practitioners treat model-building code with the same testing rigor as traditional production code. Concretely, ML practitioners and Dev(ML)-ops toolsmiths can:
\begin{itemize}
    \item Introduce property-based or metamorphic tests to validate key invariants, for instance, consistent output shapes across inputs and reproducibility under fixed random seeds.
    \item Enforce a minimum  (say 80\%) statement coverage for data preprocessing and model code before allowing merges. Such checks can be integrated in CI pipelines.
    \item Implement regular drift testing: periodically (e.g., nightly), re-run the model on a frozen validation dataset and compare key performance metrics (e.g., accuracy, loss) against historical values. Triggered alerts can then be attributed to test coverages across Data and Model components, along with high-churn development phases, as mentioned previously. 
\end{itemize}

This structured approach ensures that untested or poorly tested model logic is minimized, improving both code reliability and model performance stability.\\

\textbf{4. Isolate experiments behind feature flags and short-lived branches.} Our results for RQ1 and RQ4 indicate that, in ML systems, high churn and exploratory work is inevitable. It is possible that half implemented ideas (as indicated by code contaminated with large amounts of SATD) might seep into the main codebase by junior developers or during a hurried development sprint. Safeguarding against such experimental features seeping into the main developmental branch is essential. Practitioners can use runtime flags (for instance, Hydra\footnote{Last accessed Apr 27, 2025 \url{https://hydra.cc/docs/advanced/hydra-command-line-flags/}} or LaunchDarkly\footnote{last accessed Apr 17, 2025 \url{https://launchdarkly.com}}) so researchers and practitioners can ship prototypes without altering the production path. Furthermore, Dev(or ML)-ops practitioners can enable CI rigs such that branches older than `N' weeks without merges are auto-archived; flags older than one release trigger a ``clean-up'' review. This fortifies the gates to confine debt into explicitly experimental gated areas, and prevents accidental seepage into the main developmental branch. Empirical studies of feature toggle practices~\cite{mahdavi2021software} emphasize the need for systematic cleanup routines and lifecycle management, corroborating with our call to prevent long-lived flags from becoming a persistent source of technical debt.\\

\textbf{5. Know how much you debt owe, and don't suffer later: Implement debt heat-maps within the pull-request review workflow.}
Our analysis of debt survival indicates that debt can stay for a long time in both traditional and ML software. Technical debt can be a neglected area, and just like financial debt, not knowing how much you owe can be catastrophic to systems health. We recommend implementing SATD heat-maps visualizations within CI systems such as Jenkins\footnote{last accessed Apr 17, 2025 \url{https://www.jenkins.io}}. This heat-map can visualize the output of an automated SATD detection tool such as the one we used in our research~\cite{liu2018satd}. By visualizing the code diff with intuitive colour-coded overlays, reviewers are immediately alerted to potential debt-prone regions—for example, large green-highlighted sections indicating extensive code modifications in low-complexity areas where SATD has been flagged. This visual augmentation of pull requests can also draw attention to structural weak points that might otherwise escape scrutiny. Even though payment of debt is done in the future, at least the debt-riddled commits and LOCs are not ``forgotten'' in the long run. Recently, tools like DebViz~\cite{li2023debtviz} provision identifying, measuring, visualizing, and monitoring SATD, answering to this call. 

Moreover, \textbf{setting Service-Level Objectives (SLOs)~\cite{hidalgo2020decision} for debt comments} can help in the long run. For instance, adding a bot that pings the author if an SATD comment is still present two weeks after the introduction or past a release can prevent long-term accumulation of debt. Public dashboards showing open-vs-resolved counts might also create gentle social pressure and keep removal latency low.

\textbf{6. Integrate ``debt labs'' into ML curricula.}
By educating the next generation of ML practitioners about the potential pitfalls and long-term consequences of technical debt, ML educators can contribute to improving the overall quality and maintainability of ML systems. This implication aligns with several pedagogical studies~\cite{tonin2017effects,crespo2021carrot} focused on reducing technical debt among computer science students.
Case studies and real-world examples (for instance, students must (i) identify SATD in an open-source ML repository, (ii) write a remediation plan, and (iii) implement one fix. This practice grounds theory in visible, real-world debt and builds habits that transfer to industry) should be included to demonstrate the impact of technical debt on ML projects, fostering a deep understanding of the consequences of neglecting technical debt. Given that requirement debt is most prevalent (RQ2), and the model building side accumulates the maximum amount of debt (RQ3), ML educators may wish to provide special focus on such areas. This approach can help the next generation of ML practitioners to be better equipped to manage technical debt effectively.

\section{Future Work}\label{sec.future_work}
In this section, we discuss the future work that can be inspired by our research.

\begin{itemize}
    
    \item \textbf{Examining TD (SATD) in Other Aspects of ML Development.} In RQ3, we label 611 ML Self-Admitted Technical Debt (SATD) instances onto the ML components proposed by Amershi et al~\cite{amershi2019software}. Future research may consider examining Technical Debt (TD) in ML components that are not mentioned in Amershi’s pipeline. This could involve supporting tasks like model monitoring, maintenance, and evolution, ethics and fairness evaluation, or the user feedback loop. Evaluating SATD in ML continuous integration/ continuous delivery (CI/CD) tasks like model/data versioning, EOL (end of life planning), or rollback and redeployments, etc. could also be a valuable area for the research community. Overall, future work could investigate TD's impact on the performance and robustness of ML models, as well as the overall system's reliability, security, and maintainability.

    \item \textbf{Automated TD Prioritization Through SATD Identification.} A potential avenue for future research is developing automated methods to prioritize SATD in ML-based software projects. These methods could factor in various aspects of TD, such as severity, potential consequences specific to ML, and the cost of addressing ML debt, debt arising from data handling, inconsistencies in ML models, and other integral components of ML software development. Automated approaches could assist developers and project managers in efficiently managing SATD. This could involve quickly identifying areas of concern and allocating resources to minimize negative impacts of accumulated debt on the system's quality, maintainability, and performance.

    \item \textbf{Team Dynamics and Development Processes.} Investigating the relationship between team dynamics, development processes, and the introduction of SATD in ML-based software projects can provide valuable insights into factors contributing to the accumulation of technical debt. Future research could explore the influence of factors like team size, experience, communication patterns, and development methodologies on the prevalence and types of SATD in ML-based software systems. Understanding these relationships might lead to strategies and best practices for minimizing SATD during the software development lifecycle.

    \item \textbf{Guidelines and Best Practices for Managing ML SATD.} A crucial area for future work is proposing actionable guidelines and best practices for managing and mitigating SATD in ML-based software projects. These guidelines could help developers and project managers make informed decisions when dealing with trade-offs and competing priorities during ML software development. They could focus on various aspects of SATD, including prevention, detection, prioritization, and resolution, tailored to the unique characteristics and challenges of ML-based software systems.

    \item \textbf{Leveraging ML for Detecting ML-specific SATD.} Existing SATD detection tools mainly target non-ML software. Due to the distinct nature of ML, there might be a need for ML-specific SATD detection tools that are more effective in identifying ML-related debt. Developing such tools in future research could facilitate efficient identification, tracking, and resolution of SATD throughout ML development, covering a wide range of ML tasks~\cite{amershi2019software}.

    \item \textbf{Evaluating the Impact of SATD on Model Performance.}
    Since ML-based systems contain a ML-pipeline which produces an ML model, debt in ML-based systems may impact the performance of ML models. For instance, debt in data-preprocessing or model building components may lead to a model with poor performance. Future research could explore the impact of technical debt on the model performance metrics, such as accuracy, precision, recall, and F1-score, to help ML practitioners prioritize addressing technical debt that has the most significant impact on their models.
 
    \item \textbf{Evaluation of SATD Pre-Rebasing.} Our research primarily examines the spread of SATD within the master branch. However, future studies might consider investigating the behavior of SATD across various branches. Given that change evaluation across branches is a burgeoning field of research~\cite{bhatia2023towards}, examining SATD dynamics in this context could be valuable in understanding debt.
\end{itemize}
 
\section{Conclusion}\label{sec.conclude}
Developers make suboptimal decisions and introduce technical debt while implementing ML software. Although prior studies have investigated SATD in non-ML software, the nature of SATD in ML code can be expected to be much different from non-ML code.

We conclude our study by elucidating the key contributions of our study, underscoring its significance in enhancing the existing body of research in Self-Admitted Technical Debt (SATD) and Software Engineering for Machine Learning (SE4ML).

For \textbf{RQ1}, we find that Machine Learning (ML) code accumulates twice the debt in comparison to non-ML code. While Sculley et al.~\cite{Sculley} previously hypothesized the potential for debt to cause maintenance challenges in ML software, ours is the first study to provide empirical evidence for the pronounced prevalence of debt in ML code. Notably, our finding that 2\% of non-ML code has SATD augments findings from prior SATD studies on non-ML software. Our results align and further contextualize the research by Potdar and Shihab~\cite{potdar2014exploratory}, who ascertained that debt permeates in 2.7\% to 31\% of files, 0.4\% to 3.3\% at a class level, and 0.3\% to 2.6\% at a function level. Our methodology quantifies debt based on the percentage of comments marked as SATD, facilitating a comparative analysis between ML and non-ML comments.

In \textbf{RQ2}, our qualitative efforts extend the prior taxonomy of SATD. The prior taxonomy, by Bavota and Russo~\cite{bavota2016large}, proposed seven years ago, necessitates augmentation to cater to the ML-dominated software development landscape that is more relevant today. In our qualitative analysis of SATD in ML code, we identified two new ML-specific SATD categories: ``\textit{inadequate testing}" and ``\textit{configuration}'' debts. We believe that these categories emerged due to the unique challenges of ML pipeline development, which involves frequent experimentation for model optimization and management of hyperparameters and data preprocessing configurations. These findings offer a clearer view of the technical debts present in today's ML-centric software development.

In \textbf{RQ3}, we analyzed SATD within different stages of Amershi et al.'s ML pipeline~\cite{amershi2019software}. This not only categorizes the debt but also pinpoints the stages in the ML pipeline where technical debt is more likely to occur, offering a structured approach to address ML debts.

For \textbf{RQ4}, we examined how quickly ML and non-ML debts are introduced and subsequently resolved. Our data shows that ML debts are introduced by 140 times and removed by 3.7 times more rapidly than non-ML debts. This aligns with the insights from Sculley et al.~\cite{Sculley}, which underscore the critical nature of proficient debt management in ML development. Our results complement existing research on SATD survival in non-ML software. Bavota and Russo~\cite{bavota2016large} reported debt survival in terms of commits and reported that SATD survives for a median of 266 commits, yet did not provide time-related results. Conversely, Maldonado et al.~\cite{maldonado2017empirical} indicated that SATD survives between 18.2-172.8 days, averaging at 82-613.2 days. While Maldonado et al. presented average timelines for debt survival, we employed survival analysis using the Kaplan-Meier metric. This approach offers more robust insights than raw averages because it accounts for censored data, providing a comprehensive view of life expectancy of SATD.

In \textbf{RQ5}, using a ML model interpretation, we delve into the characteristics of long-lasting debts in ML code. While past research briefly touched upon debt duration, to our knowledge, our study is the first to uncover an understanding of long-lasting debts, setting it apart from both ML and non-ML research in the area. Moreover, we use software metrics in the context of SATD. While earlier studies such as \cite{yan2018automating} utilized software metrics to detect SATD using ML models, our approach leverages these metrics to interpret the specific traits of long-lasting debt. Our findings indicate that files with lower complexity are more likely to permeate debt until a long time, suggesting developers should give heightened attention to such potentially overlooked small files.

Taken together, our results are better explained by development churn—rapid, iterative changes to models, datasets, and configurations that lead to frequent SATD introduction followed by prompt cleanup. This lens reconciles the faster addition/removal observed in \textbf{RQ4} with the higher overall prevalence in \textbf{RQ1}: high-velocity iteration inflate short-lived debt while removals keep pace. It also contextualizes \textbf{RQ5}: the debts that persist despite churn tend to reside in low-complexity files. 

Our results provide insights to practitioners towards better management of technical debt and call for more research work to improve our understanding of technical debt occurring in ML-based systems.


\balance
\bibliographystyle{IEEEtran}
\bibliography{references}
\end{document}